\documentclass[10pt,a4paper,twoside,final]{article}
\usepackage[switch]{lineno}

\usepackage[a4paper,
		top=2.5cm,
		bottom=1cm,
		textwidth=183mm,
		includefoot,
		foot=1cm,
		head=2cm]{geometry}

\usepackage{amsmath,amsfonts, amssymb, amsthm}

\usepackage{mathrsfs}

\usepackage{sidecap}

\usepackage{xcolor, framed}
\definecolor{shadecolor}{rgb}{0.9, 0.9, 0.9}

\usepackage[]{graphicx}

\usepackage{multicol, multirow}

\usepackage{tocloft}

\usepackage[font=small,labelfont=bf,sc]{caption}
\usepackage{mdframed}
\usepackage{calc}
\newlength{\mylength}
\newlength{\figurewidth}
\newsavebox\myboxA
\newsavebox\myboxB
\newlength\mylenA
\newcommand*\widebar[2][0.8]{%
    \sbox{\myboxA}{$\m@th#2$}%
    \setbox\myboxB\null
    \ht\myboxB=\ht\myboxA%
    \dp\myboxB=\dp\myboxA%
    \wd\myboxB=#1\wd\myboxA
    \sbox\myboxB{$\m@th\overline{\copy\myboxB}$}
    \setlength\mylenA{\the\wd\myboxA}
    \addtolength\mylenA{-\the\wd\myboxB}%
    \ifdim\wd\myboxB<\wd\myboxA%
       \rlap{\hskip 0.5\mylenA\usebox\myboxB}{\usebox\myboxA}%
    \else
        \hskip -0.5\mylenA\rlap{\usebox\myboxA}{\hskip 0.5\mylenA\usebox\myboxB}%
    \fi}

\usepackage[hyperref,backend=biber,style=nature, url=false, doi = false, isbn=false, doi=false, maxbibnames=2, date=year]{biblatex}
\AtEveryBibitem{
 \ifentrytype{book}{}{
  \clearlist{publisher}
  \clearname{editor}
  }
}

\usepackage[protrusion=true]{microtype}

\usepackage[english]{babel}
\usepackage{hyperref}
\usepackage{textcomp}  
\usepackage{color}
\usepackage[normalem]{ulem} 
\usepackage{csquotes}  
\usepackage{booktabs}  
\usepackage{dcolumn}

\usepackage{sistyle}
\newcommand*{\micro}{\ensureupmath{\mbox{\textmu}}}
\newcommand*{\mum}{\,\ensureupmath{\micro m}}
\newcommand*{\mm}{\,\ensureupmath{mm}}
\newcommand*{\cm}{\,\ensureupmath{cm}}
\newcommand*{\msec}{\,\ensureupmath{ms}}
\renewcommand*{\sec}{\,\ensureupmath{s}}
\newcommand*{\Hz}{\,\ensureupmath{Hz}}

\DeclareRobustCommand{\thinskip}{\hskip 0.16667em\relax}
\def\emdash{---}
\def\dash#1#2{\unskip#1\thinskip#2\thinskip\ignorespaces}
\def\Dash{\dash\nobreak\emdash}
\def\Ldash{\dash\empty{\hbox{\emdash}\nobreak}}
\def\Rdash{\dash\nobreak\emdash}

\newcolumntype{L}[1]{>{\raggedright\let\newline\\\arraybackslash\hspace{0pt}}m{#1}}
\newcolumntype{C}[1]{>{\centering\let\newline\\\arraybackslash\hspace{0pt}}m{#1}}
\newcolumntype{R}[1]{>{\raggedleft\let\newline\\\arraybackslash\hspace{0pt}}m{#1}}

\newcommand{\changed}[1]{\textcolor{black}{#1}}
\newcommand{\changedd}[1]{\textcolor{black}{#1}}
\newcommand{\removed}[1]{}

\title{Sampling effects and measurement overlap can bias the inference of neuronal avalanches}

\author{Joao Pinheiro Neto$^1$, F.~Paul Spitzner$^1$ and Viola Priesemann$^{1,2}$ \\ {\small $^1$Max Planck Institute for Dynamics and Self-Organization, G\"ottingen} \\ {\small $^2$Bernstein Center for Computational Neuroscience, G\"ottingen}}
\date{ \normalsize Preprint, \today}

\begin{document}
\twocolumn[
  \begin{@twocolumnfalse}

   {\Huge  \noindent Sampling effects and measurement overlap can bias the inference of neuronal avalanches} \\

   {\large \noindent Joao Pinheiro Neto$^{1,\dagger}$, F.~Paul Spitzner$^{1,\dagger}$ and Viola Priesemann$^{1,2,3,*}$} \\

   $^1$Max Planck Institute for Dynamics and Self-Organization, G\"ottingen, Germany \\
   $^2$Bernstein Center for Computational Neuroscience, G\"ottingen, Germany\\
   $^3$Georg-August University G\"ottingen, G\"ottingen, Germany\\
   $^\dagger$ J.P.N.~and F.P.S.~contributed equally.\\
   $^*$viola.priesemann@ds.mpg.de

\section*{Abstract}
\noindent
To date, it is still impossible to sample the entire mammalian brain with single-neuron precision.
This forces one to either use spikes (focusing on few neurons)
or to use coarse-sampled activity
(averaging over many neurons, e.g.~LFP).
Naturally, the sampling technique impacts inference about collective properties.
Here, we emulate both sampling techniques on a simple spiking model to quantify how
they alter observed correlations and signatures of criticality.
We \changedd{describe} a general effect:
when the inter-electrode distance is small, electrodes sample overlapping regions in space, which increases the correlation between the signals.
For coarse-sampled activity, this can produce power-law distributions even for non-critical systems.
In contrast, spike recordings do not suffer this \changedd{particular} bias and underlying dynamics can be identified.
This may resolve why coarse measures and spikes have produced contradicting results in the past.

\tableofcontents

\end{@twocolumnfalse}
]


\section*{Author Summary}

The criticality hypothesis associates functional benefits with neuronal systems that operate in a dynamic state at a critical point.
A common way to probe the dynamic state of a neuronal systems is measuring characteristics of
so-called avalanches \Dash distinct cascades of neuronal activity that are separated in time.
For example, the probability distribution of the avalanche size will resemble a power-law if a neuronal system is critical.
Thus, power-law distributions have become a common indicator for critical dynamics.

Here, we use simple models and numeric simulations to show that not only the dynamic state of a system has an impact on avalanche distributions.
Also aspects that are only related to the sampling of the system (\changed{such as} inter-electrode distance) or the way avalanches are calculated (such as thresholding and time binning) can shape avalanche distributions.
On a mechanistic level we find that, if electrodes record spatially overlapping regions, the signals of electrodes may be \textit{spuriously} correlated; multiple electrodes might pick up activity from the same neuron.
Subsequently, when avalanches are inferred, such a measurement overlap can produce power-law distributions even if the underlying system is not critical.



\section{Introduction}
For more than two decades, it has been argued that the cortex might operate  at a critical point~\cite{Beggs2003,Dunkelmann1994,Beggs2008,Munoz2018a,Cocchi2017,Plenz2014,zeraati_self-organization_2020}.
The criticality hypothesis states that by operating at a critical point, neuronal networks could benefit from optimal information-processing properties.
Properties maximized at criticality include the correlation length~\cite{Sethna2006}, the autocorrelation time~\cite{Plenz2014}, the dynamic range~\cite{Kinouchi2006, Zierenberg2019a} and the richness of spatio-temporal patterns~\cite{Haldeman2005, Tkacik2015}.

Evidence for criticality in the brain often derives from measurements of \textit{neuronal avalanches}.
Neuronal avalanches are cascades of neuronal activity that spread in space and time.
If a system is critical, the probability distribution of avalanche size $p(S)$ follows a power law $p(S)\sim S^{-\alpha}$~\cite{Sethna2001,Sethna2006}.
Such power-law distributions have been observed repeatedly in experiments since they were first reported by Beggs~\&~Plenz in 2003~\cite{Beggs2003}.

However, not all experiments have produced power laws and the criticality hypothesis remains controversial.
It turns out that results for cortical recordings \textit{in vivo} differ systematically:

Studies that use what we here call \textit{coarse-sampled} activity typically produce power-law distributions~\cite{Beggs2003,Gireesh2008,Petermann2009,Dehghani2012, Clawson2017,Ribeiro2010,Shriki2013,Arviv2015,Palva2013,Tagliazucchi2012,Ponce-Alvarez2018}.
In contrast,
studies that use \textit{sub-sampled} activity
typically do not~\cite{Priesemann2014, Dehghani2012, Bedard2006, Ribeiro2014, Wilting2018, Wilting2018b}.
Coarse-sampled activity include LFP, M/EEG, fMRI and potentially calcium imaging, while sub-sampled activity is front-most spike recordings.
We hypothesize that the apparent contradiction between coarse-sampled (LFP-like) data and sub-sampled (spike) data can be explained by the differences in the recording and analysis procedures.

In general, the analysis of neuronal avalanches is not straightforward.
In order to obtain avalanches, one needs to define discrete events.
While spikes are discrete events by nature, a coarse-sampled signal has to be converted into a binary form.
This conversion hinges on thresholding the signal, which can be problematic~\cite{Font-Clos2015a,Laurson2009,Villegas2019,Porta2018}.
Furthermore, events have to be grouped into avalanches, and this grouping is typically not unique~\cite{Priesemann2014}.
As a result, avalanche-size distributions depend on the choice of the threshold and temporal binning~\cite{Beggs2003,Klaus2011}.

In this work, we show how thresholding and temporal binning interact with a \changedd{commonly ignored effect~\cite{Yu2014,Dehghani2012}}.
Under coarse-sampling,
neighboring electrodes may share the same field-of-view. This creates a distance-dependent \textit{measurement overlap}
so that the activity that is recorded at different electrodes may show \textit{spurious correlations}, even if the underlying spiking activity is fully uncorrelated.
We show that the inter-electrode distance may therefore impact avalanche-size distributions more severely than the underlying neuronal activity.

In this numeric study, we explore the role of the recording and analysis procedures on a locally-connected network of simple binary neurons.
Focusing on avalanche distributions, we compare apparent signs of criticality under sub-sampling versus coarse-sampling.
To that end, we vary the distance to criticality of the underlying system over a wide range, from uncorrelated (Poisson) to highly-correlated (critical) dynamics. We then employ a typical analysis pipeline to derive signatures of criticality and study how results depend on electrode distance and temporal binning.

\section{Results}

The aim of this study is to understand \textit{how the sampling of neural activity} affects the inference of the underlying collective dynamics.
This requires us to be able to precisely set the underlying dynamics.
Therefore, we use the established branching model~\cite{Harris1963}, which neglects many biophysical details, but it allows us to precisely tune the dynamics and to set the distance to criticality.

To study sampling effects, we use a two-level setup inspired by~\cite{Yu2014}:
an underlying network model, on which activity is then \textit{sampled} with a grid of $8 \times 8$ virtual electrodes.
Where possible, parameters of the model, the sampling and the analysis are motivated by values from experiments (see Methods).

In order to evaluate sampling effects, we want to \textit{precisely} set the underlying dynamics.
The branching model meets this requirement and is well understood analytically~\cite{Haldeman2005,Harris1963,Yu2014,Zierenberg2019, Wilting2018}.
Inspired by biological neuronal networks, we simulate the branching dynamics on a 2D topology with $N_{\rm N}=160\,000$ neurons where each neuron is connected to $K \approx 1000$ local neighbors.
To emphasize the locality, the synaptic strength of connections decays with the distance $d_{\rm N}$ between neurons.
For a detailed comparison with different topologies, see the Supplemental Information (Fig~\ref{fig:supp_topologies}).

\begin{figure}[!t]
\includegraphics[width=\columnwidth]{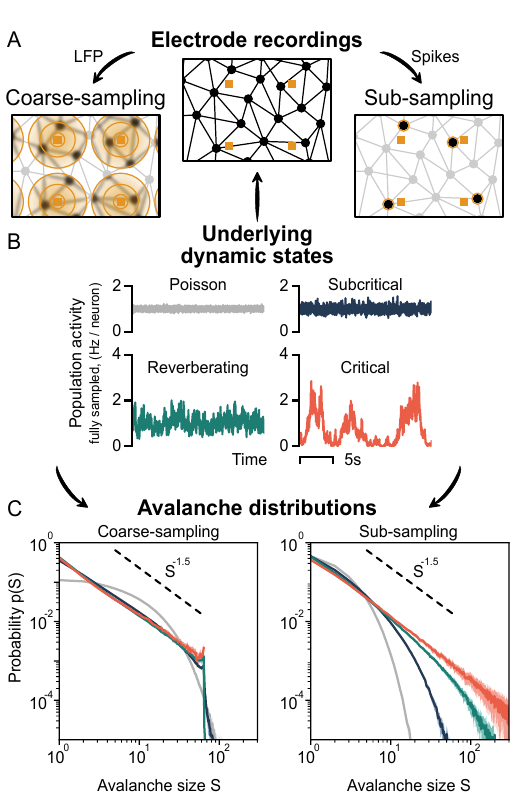}
\caption{\textbf{Sampling affects the assessment of dynamic states from neuronal avalanches.}
\textbf{A:} Representation of the sampling process of neurons (black circles) using  electrodes (orange squares). Under coarse-sampling (e.g.~LFP), activity is measured as a weighted average in the electrode's vicinity. Under sub-sampling (spikes), activity is measured from few individual neurons.
\textbf{B:} Fully sampled population activity of the neuronal network, for states with varying intrinsic timescales $\tau$: Poisson ($\hat{\tau}_{\rm psn} \approx 0 \msec$), subcritical ($\hat{\tau}_{\rm sub} \approx 19 \msec$), reverberating ($\hat{\tau}_{\rm rev} \approx 98 \msec$) and critical ($\hat\tau_{\rm crit} \approx 1.6 \sec$).
\textbf{C:} Avalanche-size distribution $p(S)$ for coarse-sampled (left) and sub-sampled (right) activity. Sub-sampling allows for separating the different states, whereas coarse-sampling leads to $p(S)\sim S^{-\alpha}$ for all states except Poisson. \textbf{Parameters:}~\changed{Electrode contribution $\gamma=1$,} inter-electrode distance $d_{\rm E}=400\mum$ and time-bin size $\Delta t = 8 \msec$.}
\label{fig:sampling}
\end{figure}

\subsection{Avalanches are extracted differently under coarse-sampling and sub-sampling}
At each electrode, we sample both the spiking activity of the closest neuron (sub-sampling) and a spatially averaged signal that emulates LFP-like coarse-sampling.

Both \textit{coarse-sampling} and \textit{sub-sampling} are sketched in Fig~\ref{fig:sampling}A:
For coarse-sampling (left), the signal from
each electrode channel is composed of varying contributions (orange circles) of all surrounding neurons.
The contribution of a particular spike from neuron $i$ to electrode $k$ decays as $1/d_{ik}^\gamma$ with the neuron-to-electrode distance $d_{ik}$ \changed{and electrode contribution $\gamma =1$.}
In contrast, if spike detection is applied (Fig~\ref{fig:sampling}A, right), each electrode signal captures the spiking activity of few individual neurons (highlighted circles).

In order to focus on the key mechanistic differences between the two sampling approaches, we keep the two models as simple as possible. (This also matches the simple underlying dynamics, for which we can precisely set the distance to criticality).
However, especially for coarse-sampling, this yields a rather crude approximation:
More realistic, biophysically detailed LFP models would yield much more complex distance dependencies, which are an open field of research~\cite{Pettersen2008,Linden2011, riera_pitfalls_2012, Einevoll2013}.
\changed{Our chosen electrode-contribution of $\gamma=1$ assumes a large field of view, which implies the strongest possible measurement overlap to showcase the coarse-sampling effect.
As this is an important assumption, we consider electrodes with a smaller field of view in Sec.~\ref{sec:electrode_contributions} and provide an extended discussion in the Supplemental Information (Fig~\ref{fig:supp_gamma_gaus_comparison}).
}

\removed{
Thus, we show results for alternative distance dependencies and provide an extended discussion in the Supplemental Information.
}

\begin{figure}[t]
\includegraphics{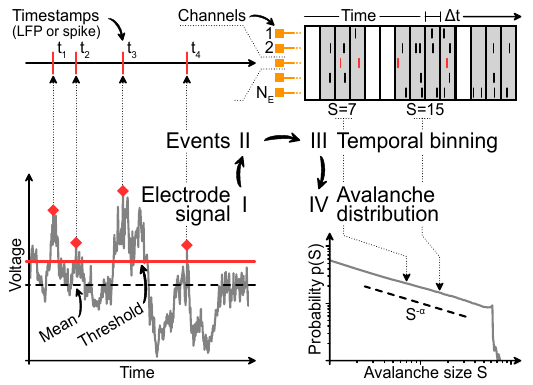}
\caption{\textbf{Analysis pipeline for avalanches from sampled data.}
\textbf{I:}~Under coarse-sampling (LFP-like), the recording is demeaned and thresholded. \textbf{II:}~The timestamps of events are extracted. Under sub-sampling (spikes), timestamps are obtained directly. \textbf{III:}~Events from all channels are binned with time-bin size $\Delta t$ and summed. The size $S$ of each neuronal avalanche is calculated. \textbf{IV:}~The probability of an avalanche size is given by the (normalized) count of its occurrences throughout the recording.
}
\label{fig:methods}
\end{figure}

To test both recording types for criticality, we apply the standard analysis that provides a probability distribution $p(S)$ of the \textit{avalanche size} $S$:
In theory, an avalanche describes a cascade of activity where individual units \Ldash here neurons \Rdash are consecutively and causally activated.
Each activation is called an event.
The avalanche size is then the total number of events in the time between the first and the last activation.
A power law in the size distribution of these avalanches is a hallmark of criticality~\cite{Plenz2014}.
In practice, the actual size of an avalanche is hard to determine because individual avalanches are not clearly separated in time;
the coarse-sampled signal is continuous-valued and describes the local population.
In order to extract binary events for the avalanche analysis (Fig~\ref{fig:methods}), the signal has to be thresholded \Dash which is not necessary for spike recordings, where binary events are inherently present as timestamps.

\subsection{The branching parameter $m$ sets the distance to criticality}
In order to compare apparent signatures of criticality with the true, underlying dynamics, we first give some intuition about the branching model.
The \textit{branching parameter} $m$ quantifies the probability of \textit{postsynaptic activations}, or in other words, how many subsequent spikes are caused (on average) by a single spike.
With increasing $m \rightarrow 1$, a single spike triggers increasingly long cascades of activity.
These cascades determine the timescale over which fluctuations occur in the population activity \Dash
this \textit{intrinsic timescale} $\tau$ describes the dynamic state of the system and its distance to criticality.

\begin{table}[b!]
  \centering
  \caption{Parameters and intrinsic timescales of dynamic states. All combinations of branching parameter $m$ and per-neuron drive $h$ result in a stationary activity of $1 \Hz$ per neuron.
  Due to the recurrent topology, it is more appropriate to consider the measured autocorrelation time $\hat{\tau}$ rather than the analytic timescale $\tau$. }
  \footnotesize
 \hspace*{-.4cm}
 \begin{tabular}{  r || c | c |  c | c }
    State name
    & $m$
    & $\hat{\tau}$ (measured)
    & $\tau=\frac{- 2 \msec}{\ln{m}}$
    & $h$
    \\
    \midrule\midrule

    Poisson
    & $ 0.0$
    & $ 0.1 \pm 0.1  \msec$
    & $   0.0 \msec$
    & $2\times10^{-3}$
    \\
    Subcritical
    & $ 0.9$
    & $ 18.96 \pm 0.09  \msec$
    & $  18.9 \msec$
    & $2\times10^{-4}$
    \\
    Reverberating
    & $0.98$
    & $  98.3 \pm 1.0  \msec$
    & $  98.9 \msec$
    & $4\times10^{-5}$
    \\
    Critical
    & $0.999$
    & $1.58 \pm 0.12 \sec$
    & $1.99 \sec$
    & $2\times10^{-6}$
    \\

  \end{tabular}
  \normalsize
  \label{tab:stateassignment}
\end{table}

The intrinsic timescale can be analytically related to the branching parameter by $\tau \sim -1/\ln{(m)}$.
As $m \to 1$, $\tau \to \infty$ and the population activity becomes \enquote{bursty}.
We illustrate this in Fig~\ref{fig:sampling}B and Table~\ref{tab:stateassignment}:
For Poisson-like dynamics ($m\approx0$), the intrinsic timescale is zero ($\hat{\tau}_{\rm psn} \approx 0 \msec$) and the activity between neurons is uncorrelated.
As the distance to criticality becomes smaller ($m\to1$), the intrinsic timescale becomes larger ($\hat{\tau}_{\rm sub} \approx 19 \msec$, $\hat{\tau}_{\rm rev} \approx 98 \msec$, $\hat{\tau}_{\rm crit} \approx 1.6 \sec $), fluctuations become stronger, and the spiking activity becomes more and more correlated in space and time.
Apart from critical dynamics, of particular interest in the above list is the \enquote{reverberating regime}:
For practical reasons, we assign a specific value of $m$ (Table~\ref{tab:stateassignment}), which represents typical values observed \textit{in vivo}~\cite{Wilting2019a, ma_cortical_2019}.
However, this choice is meant as a representation for a regime that is close-to-critical, but not directly at the critical point.
In this regime, many of the benefits of criticality emerge, while the system can maintain a safety-margin from instability~\cite{Wilting2019a}.

\subsection{Coarse-sampling can cloud differences between dynamic states}
Irrespective of the applied sampling, the inferred avalanche distribution \textit{should} represent the true dynamic state of the system.

However, under coarse-sampling (Fig~\ref{fig:sampling}C, left), the avalanche-size distributions of the subcritical, reverberating and critical state are virtually indistinguishable.
Intriguingly, all three show a power law.
The observed exponent $\alpha=1.5$ is associated with a critical branching process.
Only the uncorrelated (Poisson-like) dynamics produce a non-power-law decay of the avalanche-size distribution.

Under sub-sampling (Fig~\ref{fig:sampling}C, right), each dynamic state produces a unique avalanche-size distribution.
Only the critical state, with the longest intrinsic timescale,
produces the characteristic power law.
Even the close-to-critical, reverberating regime is clearly distinguishable and features a \enquote{subcritical decay} of $p(S)$.

\begin{figure}[!t]
\centering
\includegraphics[width=\columnwidth]{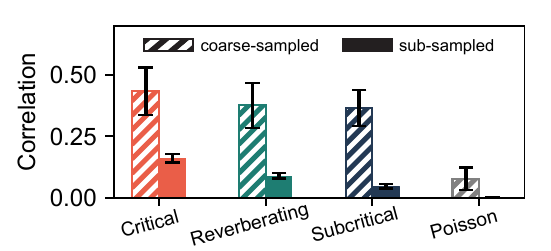}
\caption{\textbf{Coarse-sampling leads to greater correlations than sub-sampling.}
Pearson correlation coefficient between the signals of two adjacent electrodes for the different dynamic states.
Even for independent (uncorrelated) Poisson activity, measured correlations under coarse-sampling are non-zero.
 \textbf{Parameters:}~\changed{Electrode contribution $\gamma=1$,} inter-electrode distance $d_{\rm E}=400\mum$ and time-bin size $\Delta t = 8 \msec$.
}
\label{fig:corr}
\end{figure}

\begin{figure*}[!t]
\centering
\includegraphics{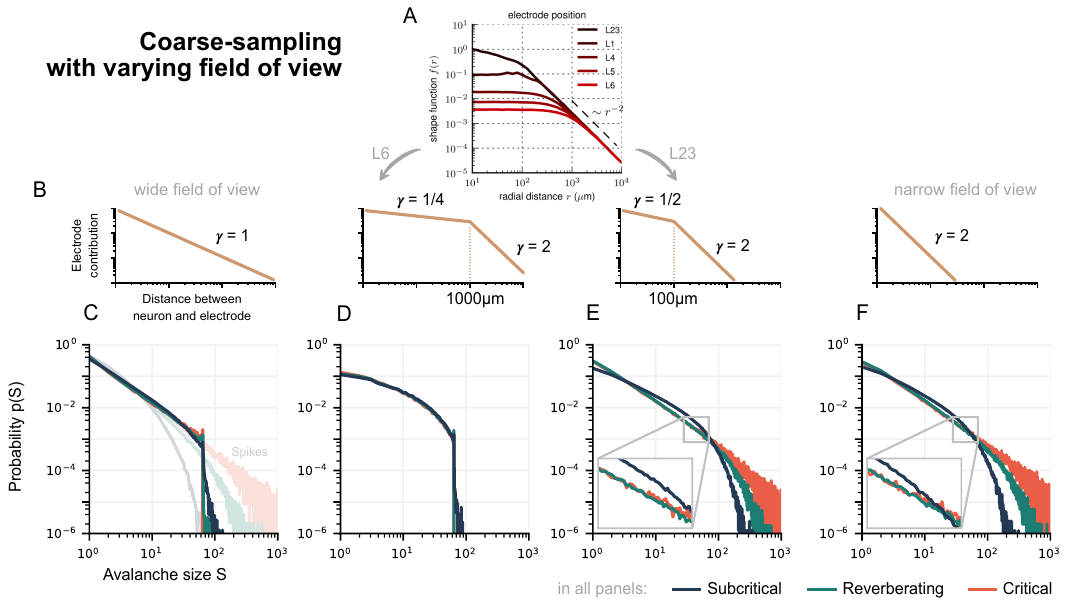}
\caption{
\changed{
\textbf{The signal of an extracellular neuronal recording depends on neuronal morphologies, tissue filtering, and other factors, which all impact the coarse-sampling effect.}
In effect, an important factor is the distance of the neuron to the electrode. Here, we show how the distance-dependence, with which a neuron's activity contributes to an electrode, determines the collapse of avalanche distributions.
\textbf{A:} Biophysically plausible distance dependence of LFP, reproduced from~\cite{Linden2011} [License will be purchased after acceptance].
\textbf{B:} Sketch of a neuron's contribution to an electrode at distance $d_{ik}$, as motivated by (A). The decay exponent $\gamma$ characterizes the field of view.
\textbf{C--F:} Avalanche-size distribution $p(S)$ for coarse-sampling with the sketched electrode contributions.
\textbf{C, D:} With a wide-field of view, distributions are hardly distinguishable between dynamic states. In contrast, for spiking activity the differences are clear (light shades in C).
\textbf{E, F:} With a narrower field of view, distributions do not fully collapse on top of each other, but differences between reverberating and critical dynamics remain hard to identify.
\textbf{Parameters:}~Inter-electrode distance $d_{\rm E}=400\mum$ and time-bin size $\Delta t = 8 \msec$. Other parameter combinations in Supplemental Figure~\ref{fig:supp_gamma_gaus_comparison}.
}
}
\label{fig:main_electrode_contrib}
\end{figure*}

\subsection{Measurement overlap causes spurious correlations}
Why are the avalanche-size distributions of different dynamic states hard to distinguish under coarse-sampling?
The answer is hidden within the cascade of steps involved in the recording and analysis procedure.
Here, we separate the impact of the involved processing steps.
Most importantly, we discuss the consequences of \textit{measurement overlap} \Dash which we identify as a key explanation for the ambiguity of the distributions under coarse-sampling.

In order to obtain discrete events from the continuous time series for the avalanche analysis, each electrode signal is filtered and thresholded, binned with a chosen time-bin size $\Delta t$ and, subsequently, the events from all channels are stacked.
This procedure is problematic because
\textbf{(i)}
electrode proximity adds spatial correlations,
\textbf{(ii)}
temporal binning adds temporal correlations,
and \textbf{(iii)}
thresholding adds various types of bias~\cite{Font-Clos2015a, Laurson2009, Villegas2019}.

As a result of the involved analysis of coarse-sampled data, spurious correlations are introduced that are not present in sub-sampled data.
We showcase this effect in Fig~\ref{fig:corr}, where the Pearson correlation coefficient between two virtual electrodes is compared for both the (thresholded and binned) coarse-sampled and sub-sampled activity.
For the same parameters and dynamic state, coarse-sampling leads to larger correlations than sub-sampling.

Depending on the \changed{sensitivity and} distance between electrodes, multiple electrodes might record activity from the same neuron.
This \textbf{measurement overlap} (or volume conduction effect) increases the spatial correlations between electrodes \Dash and because the signals from multiple electrode channels are combined in the analysis, correlations can originate from measurement overlap alone.

\changed{
\subsection{Measurement overlap depends on electrodes' field of view}
\label{sec:electrode_contributions}
The amount of measurement overlap between electrodes is determined effectively by the electrodes' field of view, thus the distance dependence with which a neuron's activity $s_{i}$ contributes to the electrode signal $V_{k}$ (Fig~\ref{fig:main_electrode_contrib}).
We consider electrode signals $V_{k}(t) = \sum_{i}^{N_{\rm N}} s_{i}(t)/d_{ik}^\gamma$, where the exponent $\gamma$ indicates how narrow ($\gamma = 2$) or wide ($\gamma = 1$) the field of view is.
Note that realistic distance dependencies are more complex and depend on many factors, such as neuron morphology and tissue filtering~\cite{Pettersen2008,Linden2011, riera_pitfalls_2012, Einevoll2013}.
}

\changed{
We find that the collapse of avalanche-size distributions from different dynamic states is strongest when the field of view is wide \Dash i.e.~if there is stronger measurement overlap. In that case, coarse-sampled distributions are hardly distinguishable (Fig~\ref{fig:main_electrode_contrib}C,~D).
For a narrow field of view, distributions are still hard to distinguish but do not fully collapse (Fig~\ref{fig:main_electrode_contrib}E,~F).
}

\changed{
In order to study the impact of inter-electrode distance and temporal binning, in the following we focus on the wide field of view ($\gamma = 1$) where the avalanche collapse is most pronounced.
}

\subsection{\changedd{The effect of inter-electrode distance}}
\begin{figure}[!t]
\includegraphics[width=\columnwidth]{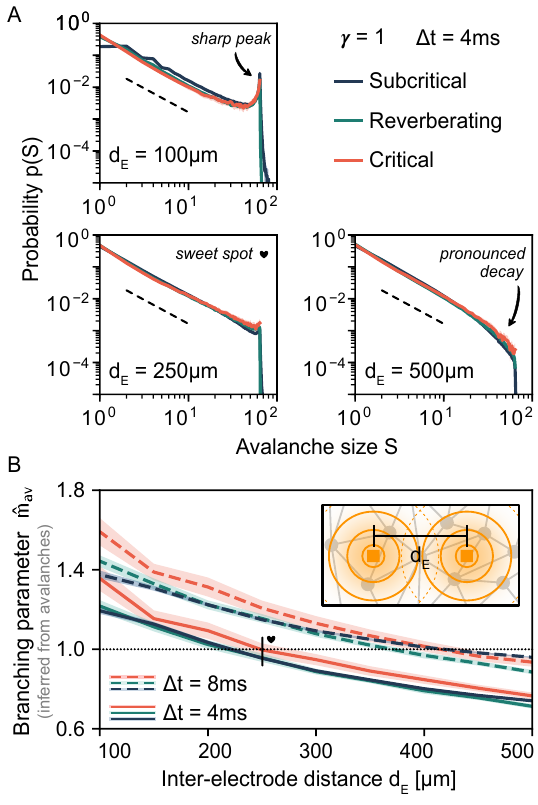}
\caption{\textbf{Under coarse-sampling, apparent dynamics depend on the inter-electrode distance $d_{\rm E}$.}
\textbf{A:}~For small distances ($d_{\rm E}=100\mum$), the avalanche-size distribution $p(S)$ indicates (apparent) supercritical dynamics:
$p(S)\sim S^{-\alpha}$ with a \textit{sharp peak} near the electrode number $N_{\rm E}=64$.
For large distances ($d_{\rm E}=500\mum$), $p(S)$ indicates subcritical dynamics:
$p(S)\sim S^{-\alpha}$ with a \textit{pronounced decay} already for $S<N_{\rm E}$.
There exists a \textit{sweet-spot} value ($d_{\rm E}=250\mum$) for which $p(S)$ indicates critical dynamics:
$p(S)\sim S^{-\alpha}$ until the the cut-off is reached at $S=N_{\rm E}$.
The particular sweet-spot value of $d_{\rm E}$ depends on time-bin size (here, $\Delta t = 4 \msec$).
As a guide to the eye, dashed lines indicate $S^{-1.5}$.
\textbf{B:}~The inferred branching parameter $\hat{m}_{\rm av}$ is also biased by $d_{\rm E}$ when estimated from neuronal avalanches. Apparent criticality ($\hat{m}_{\rm av}\approx1$, dotted line) is obtained with $d_{\rm E}=250\mum$ and $\Delta t = 4 \msec$ but also with $d_{\rm E}=400\mum$ and $\Delta t = 8 \msec$. \textbf{B, Inset:} representation of the measurement overlap between neighboring electrodes; when electrodes are placed close to each other, spurious correlations are introduced.}
\label{fig:ied}
\end{figure}

\changed{Similar to the field of view of electrodes,} avalanche-size distributions under coarse-sampling depend on the inter-electrode distance $d_{\rm E}$~(Fig~\ref{fig:ied}A).
For small inter-electrode distances, the overlap is strong. Thus, the spatial correlations are strong.
Strong correlations manifest themselves in \textit{larger} avalanches.
However, under coarse-sampling the maximal observed size $S$ of an avalanche is in general limited by the number of electrodes $N_{\rm E}$~\cite{Yu2014} (cf.~Fig~\ref{fig:supp_gamma_gaus_comparison}).
This limit due to $N_{\rm E}$ manifests as a sharp cut-off and \Ldash in combination with spurious measurement correlations due to $d_{\rm E}$ \Rdash can shape the probability distribution.
In the following, we show that these factors can be more dominant than the actual underlying dynamics.

In theory, supercritical dynamics are characterized by a \textit{sharp peak} in the avalanche distribution at $S=N_{\rm E}$.
Independent of the underlying dynamics, such a peak can originate from small electrode distances (Fig~\ref{fig:ied}A, $d_{\rm E} = 100 \mum$):
Avalanches are likely to span the small area covered by the electrode array.
Furthermore, due to strong measurement overlap, individual events of the avalanche may contribute strongly to multiple electrodes.

Subcritical dynamics are characterized by a \textit{pronounced decay} already for $S<N_{\rm E}$.
Independent of the underlying dynamics, such a decay can originate from large electrode distances (Fig~\ref{fig:ied}A,  $d_{\rm E} = 500 \mum$):
Locally propagating avalanches are unlikely to span the large area covered by the electrode array.
Furthermore, due to the weaker measurement overlap,
individual events of the avalanche may contribute strongly to one electrode (or to multiple electrodes but only weakly).

Consequently, there exists a \textit{sweet-spot} value of the inter-electrode distance $d_{\rm E}$ for which $p(S)$ appears convincingly critical (Fig~\ref{fig:ied}A,  $d_{\rm E} = 250 \mum$): a power law $p(S)\sim S^{-\alpha}$ spans all sizes up to the cut-off at $S=N_{\rm E}$.
However, the dependence on the underlying dynamic state is minimal.

\begin{figure*}[t!]
\centering
\includegraphics{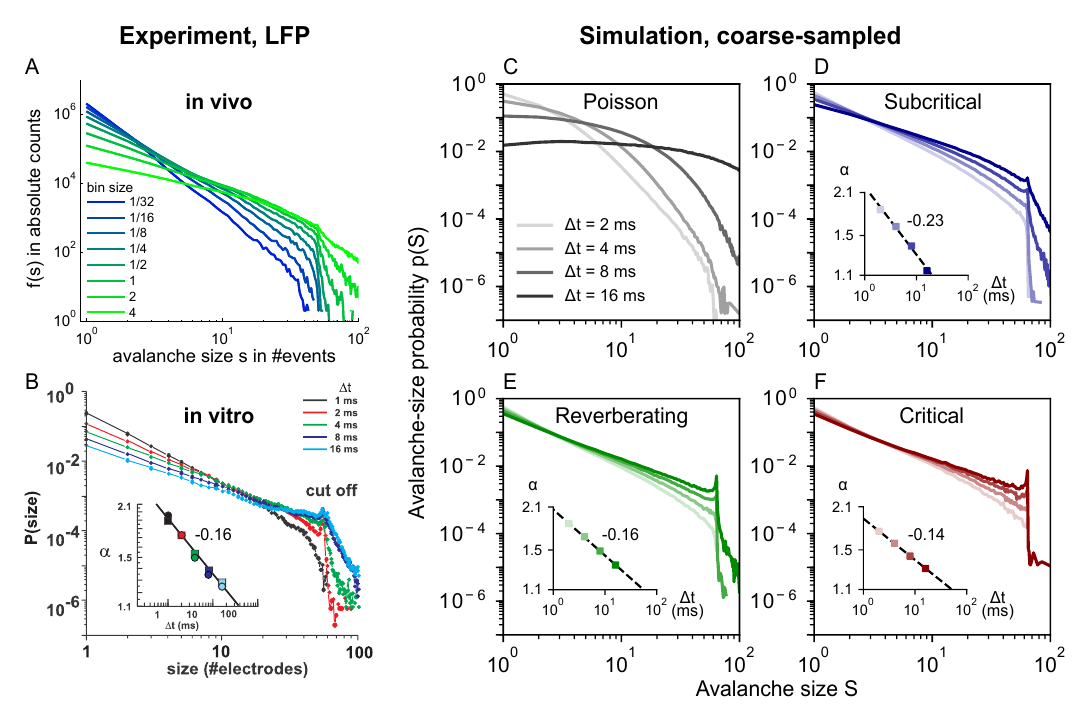}
\caption{\textbf{In vivo and in vitro avalanche-size distributions $p(S)$ from LFP depend on time-bin size $\Delta t$.}
Experimental LFP results are reproduced by many dynamics states of coarse-sampled simulations.
\textbf{A:}~Experimental \emph{in~vivo} results (LFP, human) from an array of $60$ electrodes, adapted from~\cite{Priesemann2013}.
\textbf{B:}~Experimental \emph{in~vitro} results (LFP, culture) from an array with $60$ electrodes, adapted from~\cite{Beggs2003}.
\textbf{C--F:}~Simulation results from an array of $64$ virtual electrodes and varying dynamic states, with time-bin sizes between $2 \msec \leq \Delta t \leq 16 \msec$, \changed{$\gamma=1$} and $d_{\rm E} = 400 \mum$.
Subcritical, reverberating and critical dynamics produce approximate power-law distributions with bin-size-dependent exponents~$\alpha$.
\textbf{Insets:}~\changed{Log-Log plot,} distributions are fitted to $p(S) \sim S^{-\alpha}$, \changed{fit range $S \leq 50$}. The magnitude of $\alpha$ decreases as $\Delta t^{-\beta}$ with $-\beta$ indicated next to the insets. See also Table~\ref{tab:fit_exponents}.
}
\label{fig:binsize}
\end{figure*}

\begin{figure*}[th!]
\centering
\includegraphics{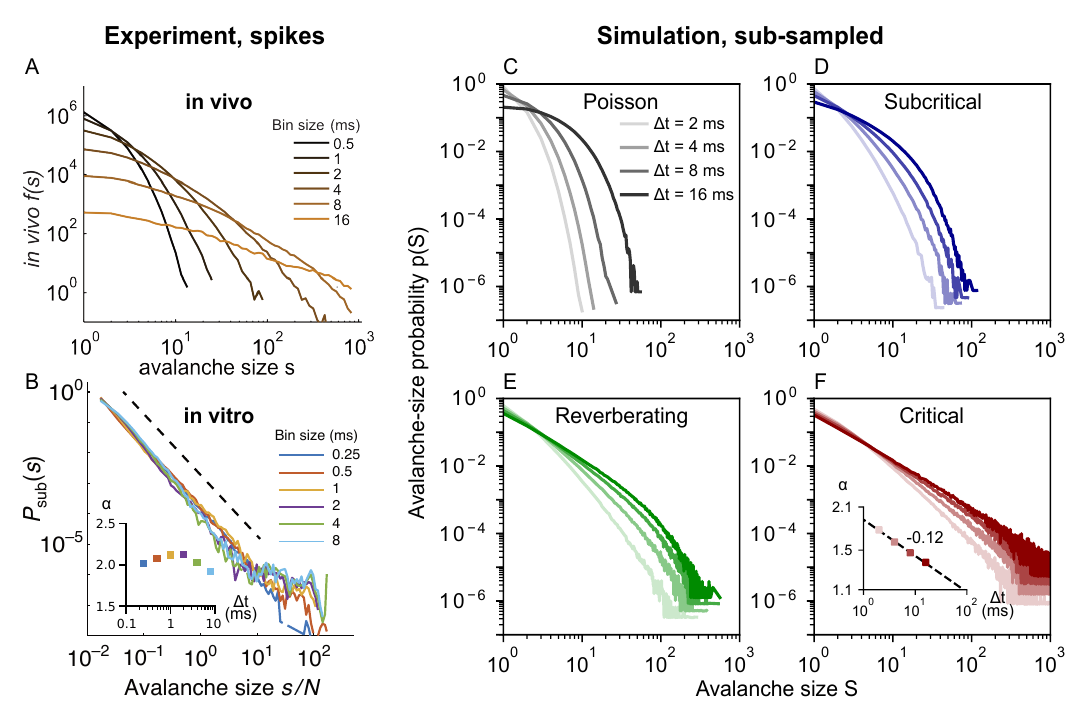}
\caption{\textbf{\emph{In~vivo} avalanche-size distributions $p(S)$ from spikes depend on time-bin size $\Delta t$.
\emph{In~vivo} results from spikes are reproduced by sub-sampled simulations of subcritical to reverberating dynamics.
}
Neither spike experiments nor sub-sampled simulations show the cut-off that is characteristic under coarse-sampling.
\textbf{A:}~Experimental \emph{in~vivo} results (spikes, awake monkey) from an array of $16$ electrodes, adapted from~\cite{Priesemann2014}.
The pronounced decay and the dependence on bin size indicate subcritical dynamics.
\textbf{B:}~Experimental \emph{in~vitro} results (spikes, culture DIV 34) from an array with $59$ electrodes, adapted from~\cite{Levina2017}.
Avalanche-size distributions are \changed{largely} independent of time-bin size and \changed{resemble} a power law over four orders of magnitude.
In combination, this indicates a separation of timescales and critical dynamics \changed{(or even super critical dynamics~\cite{plenz_self-organized_2021})}.
\changed{\textbf{B,~Inset:~} Log-Lin plot of fitted $\alpha$, fit range $s/N \leq 5$.}
\textbf{C--F:}~Simulation for sub-sampling, analogous to Fig~\ref{fig:binsize}.
Subcritical dynamics do not produce power-law distributions and are clearly distinguishable from critical dynamics.
\textbf{F:}~Only the (close-to) critical simulation produces power-law distributions.
\changed{\textbf{F,~Inset:~} Log-Log plot of fitted $\alpha$, fit range $S \leq 50$.}
In contrast to the \emph{in~vitro} culture (in B), the simulation does not feature a separation of time scales (due to external drive and stationary activity), and therefore the slope shows a systematic bin-size dependence here.
}
\label{fig:binsize_sub}
\end{figure*}

\begin{table}[b]
\centering
\caption{Fitted exponents of $\alpha \sim \Delta t^{-\beta}$.}
\begin{tabular}{lcc}
\toprule
Dynamic state& \multicolumn{2}{c}{$\beta$}\\
\cmidrule(lr){2-3}
& $d_{\rm E}=200\mum$ & $d_{\rm E}=400\mum$ \\
\midrule
in vitro (LFP)~\cite{Beggs2003}  & $0.16\pm0.01$ & \\
Critical (coarse) & $0.113\pm0.001$ & $0.141\pm0.001$\\
Reverberating (coarse) & $0.127\pm0.003$ & $0.156\pm0.002$\\
Subcritical (coarse) & $0.159\pm0.004$ & $0.231\pm0.016$\\
Critical (spikes) & $0.143\pm0.010$ & $0.123\pm0.005$\\
\bottomrule
\end{tabular}
\label{tab:fit_exponents}
\end{table}

Independently of the apparent dynamics, we observe the discussed cut-off at $S=N_{\rm E}$, which is also often seen in experiments (Fig~\ref{fig:binsize}).
Note, however, that this cut-off only occurs under coarse-sampling (see again Fig~\ref{fig:sampling}C).
When spikes are used instead (Fig~\ref{fig:binsize_sub}), the same avalanche can reach an electrode repeatedly in quick succession \Dash whereas such double-events are circumvented when thresholding at the population level.
For more details see Fig~\ref{fig:supp_gamma_gaus_comparison}.

A further signature of criticality is obtained by inferring the branching parameter. If the inference is unbiased, the inferred $\hat{m}$ matches the underlying branching parameter $m$.
We have developed a sub-sampling invariant estimator (based on the population activity inferred from spikes~\cite{Wilting2018}), but $\hat{m}$ is traditionally inferred from avalanches. Then, $\hat{m}_{\rm av}$ is defined as the average ratio of events between subsequent time bins in an avalanche, i.e.~during non-zero activity~\cite{Beggs2003, Klaus2011}.

Obtaining $\hat{m}_{\rm av}$ for different electrode distances results in a picture consistent with the one from avalanche-size distributions (Fig~\ref{fig:ied}B).
In general, the dependence on the electrode distance is stronger than the dependence on the underlying state.
At the particular value of the inter-electrode distance where $\hat{m}_{\rm av} = 1$, the distributions appear critical.
If $\hat{m}_{\rm av}<1$ ($\hat{m}_{\rm av}>1$), the distributions appear subcritical (supercritical).
Notably, the supercritical $m > 1$ corresponds to dynamics where activity increases indefinitely, which is not possible for systems of finite size and exposes $\hat{m}_{\rm av}>1$ as an inference effect.
\changed{
More precisely, in case of our simulations, $\hat{m}_{\rm av}$ suffers two sources of bias: firstly,  the coarse-sampling bias that is rooted in the preceding avalanche analysis, and secondly the estimator assumes a pure branching process without specific topology or coalescence effects~\cite{Zierenberg2019}.
}

Concluding, because the probability distributions and the inferred branching parameter share the dependence on electrode distance, a wide range of dynamic states would be consistently misclassified \Dash solely as a function of the inter-electrode distance.

\subsection{Temporal binning determines scaling exponents}
Apart from the inter-electrode distance, the choice of temporal discretization that underlies the analysis may alter avalanche-size distributions.
This \textit{time-bin size} $\Delta t$ varies from study to study and it can severely impact the observed distributions~\cite{Beggs2003,Priesemann2013,Priesemann2014,Levina2017}.
With smaller bin sizes, avalanches tend to be separated into small clusters, whereas larger bin sizes tend to \enquote{glue} subsequent avalanches together~\cite{Priesemann2014}.
Interestingly, this not only leads to larger avalanches, but specifically to $p(S)\sim S^{-\alpha}$, where the exponent $\alpha$ increases systematically with bin size~\cite{Beggs2003,Priesemann2013}.
Such a changing exponent is not expected for conventional systems that self-organize to criticality: Avalanches would be \textit{separated in time}, and $\alpha$ should be fairly bin-size invariant for a large range of $\Delta t$~\cite{Priesemann2014,Levina2017, Bak1987}.

Our coarse-sampled model reproduces these characteristic experimental results (Fig~\ref{fig:binsize}).
It also reproduces the previously reported scaling~\cite{Beggs2003} of the exponent with bin size $\alpha \sim \Delta t ^{-\beta}$ (Fig~\ref{fig:binsize}, insets).
Except for the Poisson dynamics, all the model distributions show power laws.
Moreover the distributions are strikingly similar, not just to the experimental results, but also to each other.
This emphasizes how sensitive signs of criticality are to analysis parameters:
All the shown dynamic states are consistent with the ubiquitous avalanche-size distributions that are observed in coarse-sampled experiments\changedd{~\cite{plenz_self-organized_2021} (cf.~Suppl.~Table~\ref{tab:ch2_experiments})}.

When spikes are used instead, power-law distributions only arise from critical dynamics.
For comparison with the coarse-sampled results in Fig~\ref{fig:binsize}, we show avalanche-size distributions from experimental spike recordings and sub-sampled simulations in Fig~\ref{fig:binsize_sub}.
\removed{
In this case, power laws are produced only by in vitro cultures and the simulations that are (close-to) critical.
}

In vivo spike recordings of awake animals produce distributions that feature a pronounced decay instead of power laws (Fig~\ref{fig:binsize_sub}A).
\changed{
Interestingly, spike recordings of in vitro cultures often show power-laws and, here, even little-to-no bin-size dependence, which indicates a fairly good separation of timescales (Fig~\ref{fig:binsize_sub}B). In this example, the power-law extends over several orders of magnitude, and the slope does not decrease systematically with the bin size. This indicates close-to-critical dynamics; the slight bump that represents an excess of very large avalanche, however, might also point to slight super-criticality~\cite{plenz_self-organized_2021,Levina2017}.}

\changed{
Considering our simulations of sub-sampling (Fig~\ref{fig:binsize_sub}C--F), we only observe approximate power laws if the model is (close-to) critical (Fig~\ref{fig:binsize_sub}F). Note that in critical systems, the avalanche distribution should not change with bin size, and that here the bin-size dependence of the slope is caused by the finite system size and by the non-zero spike rate, which impede a proper separation of timescales.
}
Nonetheless, in contrast to coarse-sampling, the avalanche distributions that stem from sub-sampled measures (spikes) allow us to clearly tell apart the underlying dynamic states from one another.

Overall, as our results on coarse-sampling have shown, different sources of bias \Ldash here the measurement overlap and the bin size \Rdash can perfectly outweigh each other.
For instance, smaller electrode distances (that increase correlations) can be compensated by making the time-bin size smaller (which again decreases correlations).
This was particularly evident in Fig~\ref{fig:ied}B, where increasing $d_{\rm E}$ could be outweighed by increasing $\Delta t$ in order to obtain a particular value for the branching parameter $m_{\rm av}$.
The same relationship was again visible in Fig~\ref{fig:binsize}C-F:
For the shown $d_{\rm E}=400 \mum$ (see also Fig~\ref{fig:supp_binsize} for $d_{\rm E}=200 \mum$), only $\Delta t = 8 \msec$ results in $\alpha = 1.5$ \Dash the correct exponent for the underlying dynamics.
Since the electrode distance cannot be varied in most experiments, selecting anything but the one \enquote{lucky} $\Delta t$ will cause a bias.

\section{Discussion}

When inferring collective network dynamics from partially sampled systems, it is crucial to understand how the sampling biases the measured observables.
Without this understanding, an elaborate analysis procedure \Ldash such as the one needed to study neuronal avalanches from coarse-sampled data \Rdash can result in a misclassification of the underlying dynamics.

We have shown that the analysis of neuronal avalanches based on (LFP-like) coarse-sampled data can cloud differences of avalanche distributions from systems with different spatio-temporal signatures.
These signatures derive from underlying dynamic states that, in this work, range from subcritical to critical \Dash a range over which the intrinsic timescale undergoes a hundred-fold increase.
And yet, the resulting avalanche-size distributions can be ambiguous (Fig~\ref{fig:sampling}).

The ambiguity of neuronal avalanches partially originates from spurious correlations.
We have demonstrated the generation of spurious correlations
from two sampling- and processing mechanisms: measurement overlap (due to volume conduction) and temporal binning.
Other studies found further mechanisms that can generate apparent power-law distributions by (purposely or accidentally) introducing correlations into the \textit{observed} system.
For instance, correlated input introduces temporal correlations already into the \textit{underlying} system~\cite{Priesemann2018, Touboul2015}.
Along with thresholding and low-pass frequency filtering \Ldash which add temporal correlations to the \textit{observed} system~\cite{Bedard2006, Touboul2010} \Rdash this creates a large space of variables that either depend on the system, sampling and processing, or a combination of both.

As our results focus on sampling and processing, we believe that the observed impact on avalanche-size distributions is general and model independent.
We deliberately used simple models and confirmed that our results are robust to parameter and model changes:
First, our model for coarse-sampling prioritizes simplicity over biophysical details \Ldash in order to be consistent with our simplified but well-controlled neuronal dynamics \Rdash but we checked that our results are consistent with different distance-dependencies or adding a cut-off (cf.~Figs.~\ref{fig:supp_gamma_gaus_comparison},~\ref{fig:supp_dp_parameter} and accompanying discussion).
Second, employing a more realistic topology causes no qualitative difference (Fig~\ref{fig:supp_topologies}).
Third, as a proof of concept, we investigated the impact of measurement overlap in the 2D Ising model (Fig~\ref{fig:supp_ising}).
Even in such a fundamental model a measurement overlap can bias the assessment of criticality.
Lastly, we investigated scaling relations (of avalanche size- and duration distributions) and found that under coarse-sampling, the inference is severely hindered (Fig~\ref{fig:scaling}). Under sub-sampling, scaling relations hold but with a different collapse exponent than expected for our model.
This is consistent with other recent work showing that sampling can affect the collapse exponent~\cite{carvalho_subsampled_2021}.

Despite these efforts, our work remains a mechanistic modeling study and we want to stress its limitations:
Our virtual sampling did not account for neuron morphology nor the individual neuron's connectivity profiles. As spikes are non-local events, both these aspects impact the sampling range of an electrode and the decay of e.g.~an LFP signal~\cite{Linden2011,Einevoll2013}.
Sampling also depends on effects that occur prior to recording, such as possible filtering due to extracellular tissue~\cite{Gabriel1996, Bedard2006} or filtering due to neuron morphology~\cite{Buzsaki2012, Einevoll2013}.
In particular, low-pass filtering can arise from synaptic dynamics or the propagation within dendrites~\cite{linden_intrinsic_2010}.
Clearly, as high frequencies get stripped from the signal, this could attenuate deflections of the recorded time series. Because these deflections are central to the avalanche detection, low-pass filtering could, in principle, affect avalanche statistics. However, preliminary tests showed that our main result of overlapping distributions for different dynamics states remains intact when the raw time series are low-pass filtered (cf.~Fig~\ref{fig:supp_filtering}).

Our results seemingly contradict experimental studies that demonstrate that the avalanche analysis is sensitive to pharmacological manipulations such as anesthesia~\cite{Ribeiro2010,Scott2014,Bellay2015,Fagerholm2016,fekete_critical_2018}.
Following a sufficient manipulation, a system's dynamic state will change \Dash which should be reflected by a visible difference of avalanche distributions.
We showed that under coarse-sampling, the precise dynamic state could be misclassified.
Whereas \textit{subtle} differences between the avalanche distributions from different dynamic states are indeed visible (Fig~\ref{fig:ied}), in general, they are clouded under coarse-sampling due to the measurement overlap.
However, the smaller the measurement overlap becomes (e.g.~through increasing the electrode-distance), the clearer the differences between dynamic states become (Fig~\ref{fig:supp_gamma_gaus_comparison}).
In experiments the measurement overlap is unknown; it is also a priori unknown how strong a pharmacological perturbation is (relative to the equally unknown initial dynamic state) and how much coarse-sampling affects its inference.
In modeling studies such as ours, these circumstances are well controlled \Dash providing an explanation on a mechanistic level that can now be taken into consideration (and accounted for) when analyzing experimental data.


With our results on sampling effects, we can revisit the previous literature on neuronal avalanches.
In Ref.~\cite{Ribeiro2014} Ribeiro and colleagues show that \enquote{undersampling} biases avalanche distributions near criticality. In this case, undersampling was modeled by electrodes picking up a variable number of closest neurons. Here, we separated the effect of sub-sampling (electrodes cannot record all neurons) from coarse-sampling (electrodes record multiple neurons with distance-dependent contributions) and can add to previous results:
In our model, we found that coarse-sampling clouds the differences between subcritical, reverberating, and critical dynamics;
\changedd{for $\gamma=1$}, the avalanche distributions always resemble power laws \changedd{(Fig~\ref{fig:main_electrode_contrib})}.
Because of this ambiguity, the power-law distributions obtained ubiquitously from LFP, EEG, MEG and BOLD activity should be taken as evidence of neuronal activity with spatio-temporal correlations \Dash but not necessarily of criticality proper;
the coarse-sampling might hinder such a precise classification.
In this regard, the interpretation of results from calcium imaging (which has a lower temporal resolution than electrode recordings) remains open (cf.~Table~\ref{tab:ch2_experiments} for an overview).

In contrast, a more precise classification seems possible when using spikes.
If power-law distributions are observed from (sub-sampled) spiking activity, they do point to critical dynamics.
For spiking activity, we even have mathematical tools to infer the precise underlying state in a sub-sampling-invariant manner that does not rely on avalanche distributions~\cite{Wilting2018, Wilting2019}.
However, not all spike recordings point to critical dynamics:
Whereas in vitro recordings typically do produce power-law distributions~\cite{Levina2017,Tetzlaff2010a,Friedman2012,Pasquale2008}, extracellular spike recordings from awake animals typically do not~\cite{Priesemann2014,Hahn2010a,Ribeiro2010,Dehghani2012}.

Lastly, our results might offer a solution to resolve an inconsistency between avalanche distributions that derive from spikes vs.~LFP-like sampling:
For experiments on awake animals,
spike-based studies typically indicate subcritical dynamics.
Although coarse measures typically produce power laws that indicate criticality, in this work we showed that they might cloud the difference between critical and subcritical dynamics.
Consistent with both, a brain that operates in a \textit{near}-critical regime \Ldash as opposed to a fixed dynamic state \Rdash could harness benefits associated with criticality while flexibly tuning its response properties~\cite{Fox2005, Priesemann2013, Hellyer2016, Shew2015, Simola2017, Deco2011d, Hahn2017,Tomen2014}.

\removed{
In conclusion, our results methodically separate sampling effects from the underlying dynamic state, which provides a potential solution to overcome a long-standing discrepancy concerning neuronal avalanches.
}


\section{Methods}


\begin{table*}[!t]
\centering
\caption{Values and descriptions of the model parameters.}

\begin{tabular}{p{2cm}p{2cm}p{12cm}}
\toprule
Symbol & Value & Description\\
\midrule
$\Delta t$ & $2-16\msec$ & Time-bin size (duration) for temporal binning\\
$\Theta_k$ & $3$ & Activity threshold, in units of standard deviations of the time series of electrode $k$\\
$\delta t$ & $2\msec$  & Simulation time step\\
$r$ & $1 \Hz$ & Average spike rate\\
$N_{\rm N}$ & $1.6\times10^5$ & Number of neurons\\
$d_{\rm N}$ &  $50 \mum$ & Inter-neuron distance (measured between nearest neighbors)\\
$L$         & $4 \cm$ & Linear system size\\
$\rho$      & $100 /\!\mm^{2}$  & Neuronal density\\
$K$         & $1000$ &  Average network degree (outgoing connections per neuron)\\
$d_{\rm max}$ & $1.78 \mm$ & Connection length; all neurons within $d_{\rm max}$ are connected\\
$\sigma$ & $300 \mum$ & Effective length of synaptic connections, sets the distance-dependence of the probabilities of recurrent activations\\
$N_{\rm E}$ & $8\times8$  & Number of electrodes\\
$d_{\rm E}$ & $50-500 \mum$ & Inter-electrode distance\\
$d_{\rm E}^{*}$ & $10 \mum$ & Dead-zone around each electrode (no neurons present)\\
$\gamma$ & $1$ & Decay exponent. Contributions of each spike to the coarse electrode signal scale as $V(d) \sim 1/d^\gamma$. See SI for results and discussion of different electrode contributions.\\
\bottomrule
\end{tabular}
\label{tab:parameters}
\end{table*}

\subsection{Model Details}

Our model is comprised of a two-level configuration, where a 2D network of $N_{\rm N}=160000$ spiking neurons is sampled by a square array of $N_{\rm E}=8\times8$ virtual electrodes.
Neurons are distributed randomly in space (with periodic boundary conditions)
and, on average, nearest neighbors are $d_{\rm N} = 50 {\rm \mum}$ apart.
While the model is inherently unit-less, it is more intuitive to assign some
length scale \Ldash in our case the inter-neuron distance $d_{\rm N}$ \Rdash to
set that
scale:
all other size-dependent quantities can then be expressed in terms
of the chosen $d_{\rm N}$.
For instance, the linear system size $L$ can be derived by realizing that the random placement of neurons corresponds to an ideal gas. It follows that $L=2\sqrt{N_{\rm N}} \, d_{\rm N} = 4 {\rm cm}$ for uniformly distributed neurons.
(For comparison, on a square lattice, the packing ratio would be higher and it
is easy to see that the system size would be $\sqrt{N_{\rm N}} \, d_{\rm N}$.)
Given the system size and neuron number, the overall neuronal density is
$\rho = 100 /{\rm mm}^{2}$.
With our choice of parameters, the model matches typical experimental conditions in terms of inter-neuron distance and system size (see Table~\ref{tab:parameters} for details). Whereas the apparent neuron density of $\rho = 100 /{\rm mm}^{2}$ is on the lower end of literature values~\cite{wagenaar_extremely_2006,Ivenshitz2010b}, this parameter choice avoids boundary effects that can be particularly dominant near criticality due to the long spatial correlation.
The implementation of the model in C\texttt{++}, and the python code used to analyze the data and generate the figures, are available online at \href{https://github.com/Priesemann-Group/criticalavalanches}{https://github.com/Priesemann-Group/criticalavalanches}.

\subsection{Topology}
We consider a topology that enforces \textit{local}  spreading dynamics. Every neuron is connected to all of its neighbors within a threshold distance $d_{\rm max}$.
The threshold is chosen so that on average $K=10^3$ outgoing connections are
established per neuron. We thus seek the radius $d_{\rm max}$ of a disk whose
area contains $K$ neurons. Using the already known
neuron density, we find $d_{\rm max} = \sqrt{K/ \pi \rho} \approx 1.78 \mm$.
For every established connection, the probability of a recurrent activation decreases with increasing neuron distance.
Depending on the particular distance $d_{ij}$ between the two neurons $i$ and $j$, the connection has a normalized weight
$w_{ij} = e^{-d_{ij}^2/2\sigma^2} \, / \, \Omega_{i}$ (with normalization constant $\Omega_{i} = \sum_{j'} e^{-d_{ij'}^2/2\sigma^2}$).
Our weight definition approximates the distance dependence of average synaptic strength.
The parameter $\sigma$ sets the \textit{effective} distance over which connections can form ($d_{\max}$ is an upper limit for $\sigma$ and mainly speeds up computation.)
In the limit $\sigma \to \infty$, the network is all-to-all connected.
In the limit $\sigma \to 0$, the network is completely disconnected. Therefore, the effective connection length $\sigma$ enables us to fine tune \textit{how local} the dynamic spreading of activity is. In our simulations, we choose $\sigma=6 d_{\rm N} = 300 {\rm \mum}$. Thus, the overall reach is much shorter than $d_{\rm max}$ ($\sigma \approx 0.16\,d_{\rm max}$).

\subsection{Dynamics}

To model the dynamic spreading of activity, time is discretized to a chosen simulation time step, here $\delta t=2 \msec$, which is comparable to experimental evidence on synaptic transmission~\cite{Sabatini2002}.
Our simulations run for $10^6$ time steps on an ensemble of $50$ networks for each configuration (combination of parameters and dynamic state). This corresponds to  $\sim277$ hours of recordings for each dynamic state.

The activity spreading is modeled using the dynamics of a branching process with external drive~\cite{Harris1963, Wilting2018}.
At every time step $t$, each neuron $i$ has a state $s_{i}(t)=1$ (spiking) or $0$ (quiescent).
If a neuron is spiking, it tries to activate its connected neighbors \Dash so that they will spike in the next time step.
All of these recurrent activations depend on the \textit{branching parameter} $m$:
Every attempted activation has a probability $p_{ij} = m \;w_{ij}$ to succeed.
(Note that the distance-dependent weights are normalized to $1$ but the activation probabilities are normalized to $m$.)
In addition to the possibility of being activated by its neighbors, each neuron has a probability $h$ to spike spontaneously in the next time step.
After spiking, a neuron is reset to quiescence in the next time step if it is not activated again.

Our model gives us full control over the dynamic state of the system \Dash and its distance to criticality.
The dynamic state is described by the \textit{intrinsic timescale} $\tau$.
We can analytically calculate the intrinsic timescale $\tau = -\delta t/\ln{(m)}$, where $\delta t$ is the duration of each simulated time step.
Note that $m$ \Ldash the control parameter that \textit{tunes the system} \Rdash is set on the neuron level while $\tau$ is a (collective) network property (that in turn allows us to deduce an \textit{effective} $m$).
As the system is pushed more towards criticality (by setting $m \to 1$), the intrinsic timescale diverges $\tau \to \infty$.

For consistency, we measure the intrinsic timescale during simulations.
To that end, the (fully sampled) population activity at each time step is given by the number of active neurons $A(t)=\sum_i s_i(t)$.
A linear least-squares fit of the autoregressive relation $A(t+1)  = e^{-\delta t/\tau} A(t) + N_{\rm N} h$ over the full simulated time series yields an estimate $\hat{\tau}$ that describes each particular realization.

By adjusting the branching parameter $m$ (setting the dynamic state) \textit{and} the probability for spontaneous activations $h$ (setting the drive), we control the distance to criticality \textit{and} the average stationary activity.
The activity is given by the \textit{average spike rate} $r = h/(\delta t (1-m))$ of the network.
For all simulations, we fix the rate to $r=1 {\rm Hz}$ in order to avoid rate effects when comparing different states (see Table~\ref{tab:stateassignment} for the list of parameter combinations).
Note that, due to the non-zero drive $h$ and the desired stationary activity, the model cannot be perfectly critical ($\hat{\tau} \to \infty$, see Table~\ref{tab:stateassignment}).

\subsection{Coalescence Compensation}
With our probability-based update rules, it may happen that target neurons are simultaneously activated by multiple sources.
This results in so-called \textit{coalescence effects} that are particularly strong in our model due to the local activity spreading~\cite{Zierenberg2019}.
For instance, naively setting $m=1$ (with $\sigma=300\mum$) would result in an effective (measured) $\hat{m} \approx0.98$, which has considerably different properties. Compared to e.g.~$m=0.999$ this would result in a 20-fold decrease in $\tau$.

In order to compensate these coalescence effects, we apply a simple but effective fix:
If an activation attempt is successful but the target neuron is already marked to spike in the next time step, another (quiescent) target is chosen.
Because our implementation stores all the connected target neurons as a list sorted by their distance to the source, it is easy to activate the next neuron in that list.
Thereby, the equivalent probability of the performed activation is as close to the originally attempted one as possible.

\subsection{Virtual Electrode Recordings}

Our simulations are designed to mimic sampling effects of electrodes in experimental approaches. To simulate sampling, we use the readout of $N_{\rm E} = 64$ virtual electrodes that are placed in an $8 \times 8$ grid.
Electrodes are separated by an inter-electrode distance that we specify in multiples of inter-neuron distance $d_{\rm N}$.
It is kept constant for each simulation and we study the impact of the inter-electrode distance by repeated simulations spanning electrode distances between $1 d_{\rm N} = 50 {\rm \mum}$ and $10 d_{\rm N} = 500 {\rm \mum}$.
The electrodes are modeled to be point-like objects in space that have a small dead-zone of $d_{\rm E}^{*} =  d_{\rm N} / 5 = 10 {\rm \mum}$ around their origin. Within the dead-zone, no signal can be recorded (in fact, we implement this by placing the electrodes first and the neurons second \Dash and forbid neuron placements too close to electrodes.)

Using this setup, we can apply sampling that emulates either the detection of spike times or LFP-like recordings.
To model the detection of spike times, each electrode only observes the single neuron that is closest to it.
Whenever this particular neurons spikes, the timestamp of the spike is recorded. All other neurons are neglected \Dash and the dominant sampling effect is \textit{sub-sampling}.
On the other hand, to model LFP-like recordings, each electrode integrates the spiking of all neurons in the system.
Contributions are strictly positive, matching the underlying branching dynamics (for more biophysically detailed LFP models, contributions would depend on neuron types and other factors).
The contribution of a single spike, e.g. from neuron $i$ to electrode $k$, decays as $1/d_{ik}$ with the neuron-to-electrode distance. (See the Supplemental Information, Fig~\ref{fig:supp_gamma_gaus_comparison}, for a detailed discussion of the qualitative impact
of changing the distance dependence, e.g.~to $1/d_{ik}^{2}$.)
The total signal of the electrode at time $t$ is then $V_{k}(t) = \sum_{i}^{N_{\rm N}} s_{i}(t)/d_{ik}$. Diverging electrode signals are prevented by the forbidden zone around the electrodes.
For such coarse-sampled activity, all neurons contribute to the signal and the contribution is weighted by their distance.

\subsection{Avalanches}

Taking into account all $64$ electrodes, a new avalanche starts (by definition~\cite{Beggs2003}) when
there is at least one event (spike) in a time bin \Dash given there was no event in the
previous time bin (see Fig~\ref{fig:methods}).
An avalanche ends whenever an empty bin is observed (no event over the duration of the time bin).
Hence, an avalanche persists for as long as every consecutive time bin contains
at least one event \Dash which is called the \textit{avalanche duration} $D$.
From here, it is easy to count the total number of events that were recorded across all electrodes and included time bins \Dash which is called the \textit{avalanche size} $S$.
The number of occurrences of each avalanche size (or duration) are sorted into a histogram that describes the avalanche distribution.

\subsection{Analysis of Avalanches under Coarse and Sub-sampling}
We analyze avalanche size distributions in a way that is as close to
experimental practice as possible (see Fig~\ref{fig:methods}).
From the simulations described above, we obtain two outputs from each electrode:
a) a list containing spike times of the single closest neuron
and b) a time series of the integrated signal to which all neurons
contributed.

In case of the (sub-sampled) spike times a), the spiking events are
already present in binary form.
Thus, to define a neural avalanche, the only required parameter is the size of
the time bin $\Delta t$ (for instance, we may choose $\Delta t = 4 \msec$).

In case of the (coarse-sampled) time series b), binary events need to be extracted from the continuous electrode signal.
The extraction of spike times from the continuous signal relies on a criterion to differentiate if the set of observed neurons is spiking or not \Dash which is commonly realized by applying a threshold.
(Note that now thresholding takes place on the electrode level, whereas previously, an event belonged to a single neuron.)
Here, we obtain avalanches by thresholding as follows: First, all time series are frequency filtered to $0.1 \Hz < f < 200 \Hz$. This demeans and smoothes the signal (and reflects common hardware-implemented filters of LFP recordings).
Second, the mean and standard deviation of the full time series are computed for each electrode. The mean is virtually zero due to cutting low frequencies when band-pass filtering. Each electrode's threshold is set to three standard deviations above the mean.
Third, for every positive excursion of the time series (i.e.~$V_k(t)>0$), we recorded the timestamp $t=t_{\rm max}$ of the maximum value of the excursion.
An event was defined when $V_k(t_{\rm max})$ was larger than the threshold $\Theta_k$ of three standard deviations of the (electrode-specific) time series. (Whenever the signal \textit{passes the threshold}, the timestamps of all local maxima become candidates for the event; however, only the one largest maximum between two \textit{crossings of the mean} assigns the final event-time.)
Once the continuous signal of each electrode has been mapped to binary events with timestamps, the remaining analysis steps were the same for coarse-sampled and sub-sampled data.
Last, avalanche size and duration distributions are fitted to power-laws using the powerlaw package \cite{Alstott2014}.

\subsection{Code availability}
The code used to generate and analyze the data is available online at \href{https://github.com/Priesemann-Group/criticalavalanches}{https://github.com/Priesemann-Group/criticalavalanches}.

\subsubsection*{Acknowledgments}
JPN, FPS and VP received financial support from the Max Planck Society. JPN received financial support from the Brazilian National Council for Scientific and Technological Development (CNPq) under Grant No. 206891/2014-8. VP was supported by the Deutsche Forschungsgemeinschaft (DFG, German Research Foundation) under Germany’s Excellence Strategy - EXC 2067/1- 390729940. We thank Jordi Soriano, Johannes Zierenberg and all members of our group, for valuable input. We thank Johannes Zierenberg and Bettina Royen for careful proofreading of the manuscript.

\subsubsection*{Author contributions}
J.P.N., F.P.S. and V.P. designed research; J.P.N., F.P.S. and V.P. performed research; J.P.N. and F.P.S. analyzed data; J.P.N. and F.P.S. created figures; J.P.N., F.P.S. and V.P. wrote the paper.

\subsubsection*{Competing interests}
The authors declare no competing interests.

\clearpage

\printbibliography

\nolinenumbers

\clearpage

\begin{refsection}
\section{Supplementary Information}

\renewcommand{\thefigure}{S\arabic{figure}}
\setcounter{figure}{0}
\renewcommand{\thetable}{S\arabic{table}}

\renewcommand{\figurename}{Figure}


\addtocontents{toc}{\protect\thispagestyle{empty}}
\renewcommand{\cftdot}{}
\renewcommand{\listfigurename}{List of Supplementary Figures}
\renewcommand{\listtablename}{List of Supplementary Tables}
\captionsetup[figure]{list=yes}
\captionsetup[table]{list=yes}


\subsection{Sampling bias remains under alternative topologies}

The network topology used in the main paper is local: on average, each neuron is connected to its nearest $K=10^3$ neighbors.
It is of interest to check if alternative topologies can impact the distinguishability of the underlying dynamic state under coarse-sampling.

For that, we select two additional topologies.
The first ("Orlandi") mimics the growth process of a neuronal culture.
In short, axons grow outward on a semiflexible path of limited length and have a given probability to form a synapse when they intersect the (circular) dendritic tree of another neuron.
Thereby, this topology is local without requiring distance-dependent synaptic weights (refer to \cite{Orlandi2013} for more details).
The second ("Random") implements a purely random connectivity, with each neuron being connected to $K=10^3$ neurons.
Note that this is an unrealistic setup as this topology is completely non-local.

\begin{figure}[t]
\centering
\includegraphics{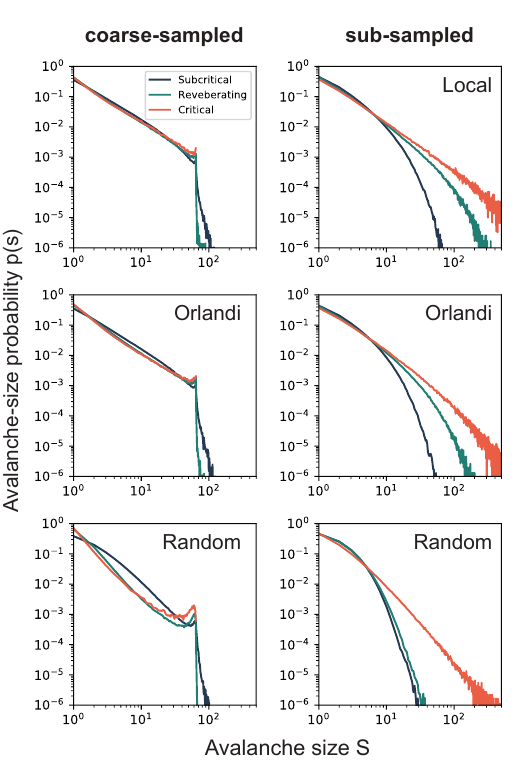}
\caption[Effect of alternative network topologies]{{\bf Effect of alternative network topologies.}
Avalanche-size probability $p(S)$ from coarse-sampled activity ({\bf left}) and sub-sampled activity ({\bf right}) for subcritical, reverberating and critical dynamics.
{\bf Top}: results for the topology used in the main paper ("Local").
{\bf Middle}: results for a topology that mimics culture growth \cite{Orlandi2013} ("Orlandi").
{\bf Bottom}: results for a random topology.
Under coarse-sampling, reverberating and critical dynamics are indistinguishable with all topologies.
Parameters: $d_{\rm E} = 400~\mum$ and $\Delta t = 8~\msec$.
}
\label{fig:supp_topologies}
\end{figure}

We find that, under coarse-sampling, reverberating and critical dynamics remain indistinguishable with the alternative topologies (Fig~\ref{fig:supp_topologies}, left).
Meanwhile, under sub-sampling,  all dynamic states are clearly distinguishable for all topologies (Fig~\ref{fig:supp_topologies}, right).

\begin{figure*}[!t]
\centering
\includegraphics[width=\textwidth]{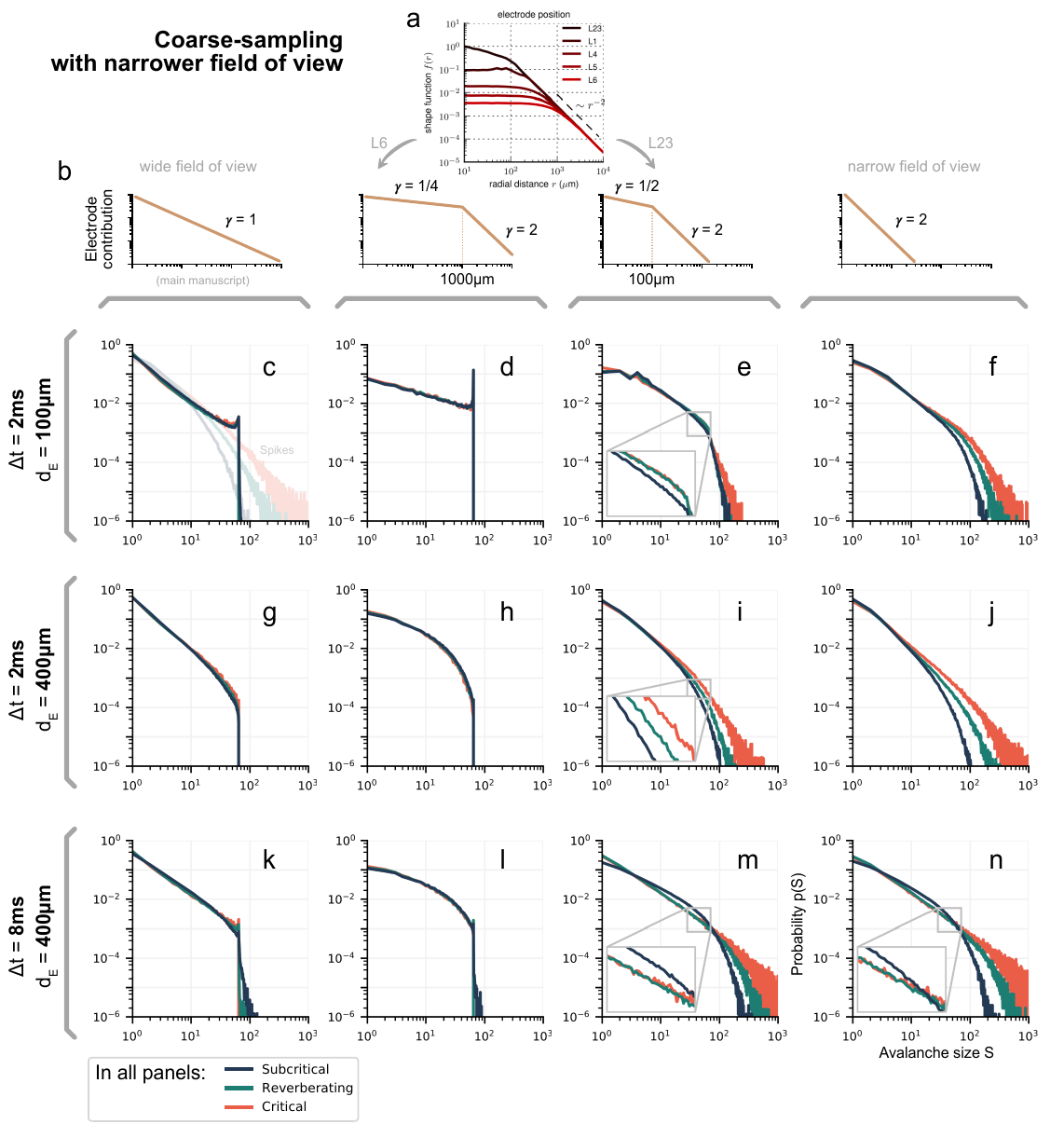}
\caption[Effect of changing the electrode contribution of a spiking neuron at distance $d$]{{\bf Effect of changing the electrode contribution of a spiking neuron at distance $d$.}
\textbf{a:} Biophysically more plausible distance dependence of LFP, reproduced from~\cite{Linden2011}.
\textbf{b:} Sketch of distance dependencies of tested effective electrode contribution that are \textit{motivated} by (A). See accompanying text.
Large decay exponents $\gamma$ correspond to a narrow field of view of the electrodes (right column).
When the transition from shallow to large exponents occurs at smaller $d$, electrodes can record fewer units and measurement overlap decreases.
Eventually, distributions become distinguishable.
As a notable side-effect, also the cut-off near $S = 64$ starts to vanish.
}
\label{fig:supp_gamma_gaus_comparison}
\end{figure*}

\begin{figure*}[!t]
\centering
\includegraphics[width=\textwidth]{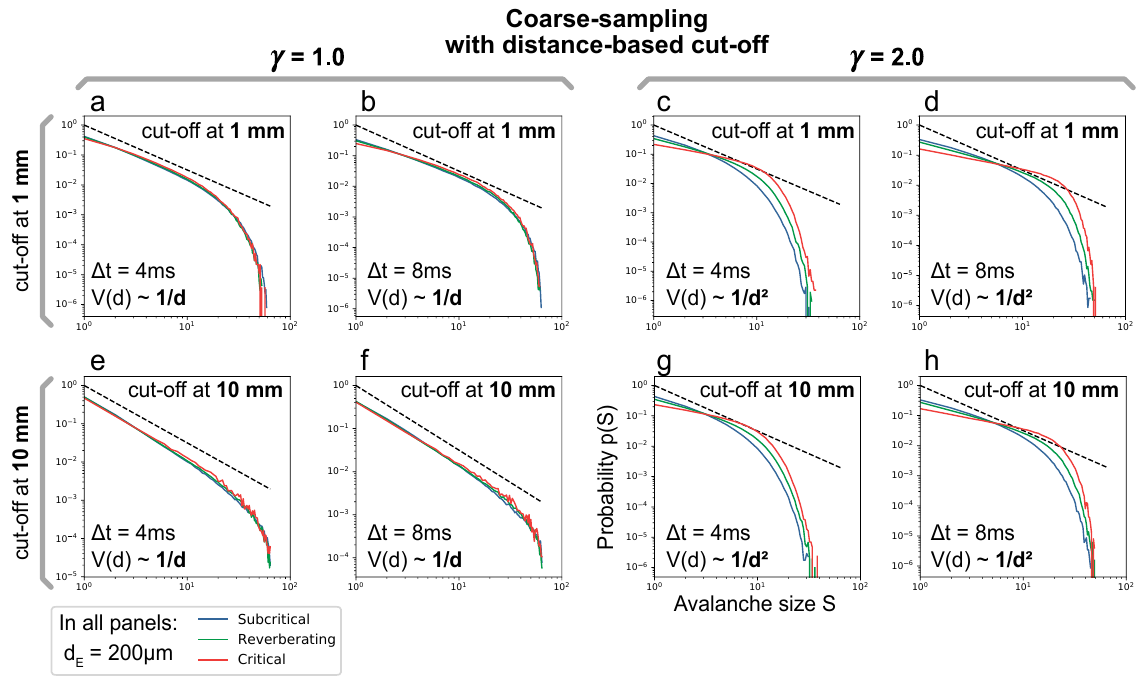}
\caption[Coarse-sampled avalanche-size distributions remain ambiguous under limited spatial reach]{{\bf Coarse-sampled avalanche-size distributions remain ambiguous under limited spatial reach.} Results from the same model of the main paper, but with a hard cut-off at 1mm (top row) and 10mm (bottom row).
Neurons farther away than the cut-off do not contribute to an electrode's signal. The effect of the cut-off stacks with the effect of the electrode decay-exponent $\gamma$.
When $\gamma = 1$, dynamic states remain indistinguishable, independent from the cut-off.
(For larger decay exponent states are distinguishable, cf.~Fig~\ref{fig:supp_gamma_gaus_comparison}.)
Notably, also for $\gamma = 1$, the shape of distributions is altered by the cut-off.
As a guide to the eye, dashed lines indicate a power law with exponent $-1.5$.
{\bf Parameters:} $\gamma = 1$ (left half) and $\gamma = 2$ (right half), $d_{\rm E} = 200 \mum$ in all panels.}
\label{fig:supp_dp_parameter}
\end{figure*}

\subsection{Influence of the electrode field-of-view}

In the main paper we considered that the contribution of a spiking neuron to the electrode signal decays with distance $d$ as $\sim 1/d$ (see Ref.~\cite{Pettersen2008,Einevoll2013}).
The precise way neuronal activity is recorded by extracellular electrodes depends on many factors such as neuronal morphology and the level of correlation between synapses~\cite{Einevoll2013, Linden2011}.
Because it is difficult to account for all relevant factors, we instead prioritized a simple, mechanistic approach.

In more detail, we are interested in the impact of a spike as a function of the distance to the electrode.
Thus, two main contributions arise: first, the transmembrane currents along the pre-synaptic neuron and the currents at the synapses and post-synaptic neurons (which do not necessarily cause further spikes).
This creates a distance-dependence of the LFP contribution that depends on the neuron's morphology and the connectivity profile.
Second, the LFP signal has a distance-dependent decay depending on whether the source corresponds to an electric monopole ($1/d$) or dipole ($1/d^2$).
The particular decay was found to depend, among other factors, on neuron type and morphology, and it effectively varies from $1/d$ near the soma to $1/d^2$ (and steeper) in the far-field limit~\cite{Pettersen2008, Linden2011, Einevoll2013}.
Together, a more realistic LFP model would need to incorporate the connectivity profile of the neurons and the signal decay (the final effect of a neuron would be a convolution of both functions).

In our simplified model, we neglect neuron morphology (because our simple binary neurons only feature a single compartment) and the connectivity profile (because our topology is homogeneous and local).
Thus, our implemented distance-dependence of the electrode signal merely serves an effective description that sensibly depends on the considered population.

As a sensitivity analysis for our effective description, we here study the impact of a varying electrode field of view (Fig~\ref{fig:supp_gamma_gaus_comparison}).
An important detail that we neglect in the main manuscript is that the $1/d$ contribution does not extend into the far-field limit (beyond $100$--$1000\mum$)~\cite{Pettersen2008,Linden2011}.
We implemented various different effective distance-dependencies, which are motivated by the work of Einevoll and colleagues, especially~Fig~2D of Ref.~\cite{Linden2011}.
Note however, that the reported shape function describes the LFP contribution as a function of a neuron receiving input, whereas, in our case, the potential describes the contribution of a spiking neuron.

In particular, we checked $\gamma=2$ (Fig~\ref{fig:supp_gamma_gaus_comparison}, right column), which represents a very narrow electrodes' field of view.
In this case, avalanche-distributions become distinguishable, but the particular shape, cut-off and the amount of overlap between states again depends on the electrode distance and time-bin size.

We also checked contributions with a shallow exponent ($\gamma \leq 1$) near the electrode, which changes into a steep exponent ($\gamma = 2$) beyond a certain distance.
For shallow exponents that reach far (transition at $1000 \mum$, Fig~\ref{fig:supp_gamma_gaus_comparison}~d,~h,~l), the distributions of all considered states overlap, as in the main manuscript.
Notably, the change in shape due to changing the inter-electrode distance is even more severe than for $\gamma = 1$.
For shallow exponents that do not reach far (transition at $100 \mum$, Fig~\ref{fig:supp_gamma_gaus_comparison}~e,~i,~m),
distributions start to be distinguishable.
However, in the usually considered range of avalanche sizes extending up to the number of electrodes ($S=64$), reverberating and critical dynamics still tend to overlap (E,~M).
Only for short time bins ($\Delta t = 2 \msec$) along large inter-electrode distances ($d_{\rm E} = 400 \mum$) are the two states distinguishable (I).

Summarizing Fig~\ref{fig:supp_gamma_gaus_comparison}, we find that a main cause of collapsing avalanche distributions is a large electrodes' field of view.
As the field of view becomes narrower, the relative contribution of the closest neurons to the electrode increases, and coarse-sampling becomes more similar to sub-sampling. The cut-off at $S\sim N_{\rm E}$ vanishes for steeper decays (large $\gamma$), and the different dynamic states become distinguishable.

For completeness, we performed further checks which follow the same reasoning.
First, instead of changing exponents, we implemented a hard cut-off, beyond which neurons cannot contribute to an electrodes' signal (Fig~\ref{fig:supp_dp_parameter}). We again found that, for cut-off values ($\geq 1 \mm$), our main finding that avalanche distributions from different dynamic states are hardly distinguishable is unaltered. For short-range cut-off values that approach the distance between neurons ($d_{\rm N} = 50 \mum$), distributions become distinguishable (not shown). This effect is similar to what we saw for increasing the decay exponent $\gamma \to 2.0$, where only individual neurons remain in the field of view of the electrode.
In both theses cases, coarse-sampling starts to observe single-unit properties and becomes similar to the case of applying spike detection (here, sub-sampling).

Second, as our electrode potential is only an effective description and cannot be directly compared to Ref.~\cite{Linden2011}, we additionally considered a hypothetical decay with exponent $\gamma = 1.5$ and repeated the analysis on different topologies (Figs.~\ref{fig:supp_gamma_d08} and~\ref{fig:supp_gamma_d04}).
In all cases, $\gamma\ge1.5$ causes the cut-off near $S=N_{\rm E}$ to vanish, and an increase of $\gamma$ weakens the coarse-sampling effect.
Note, however, that even with relatively narrow field of view, avalanche distribution from coarse-sampling still differ from those of spike analysis.

The above changes of the electrode model indicate that in real electrode recordings both effects are present, sub-sampling \textit{and} coarse-sampling. In particular, the resulting avalanche distributions change on a continuous scale where our descriptions of coarse-sampling and sub-sampling are the extremes.
Underlying dynamic states are better distinguishable when the sampling is \enquote{closer} to sampling single units instead of weighted averages \Dash and when the measurement overlap that is characteristic to coarse-sampling goes to zero.

Lastly, we want to caution about a peculiarity of $1/d^1$. From a geometric point of view, one has to consider how the number of neurons per volume element increases with distance from the electrode. In two dimensions, the number of neurons contained in a thin ring  around \changed{an electrode} scales $N \sim 2 \pi d$. If $\gamma = 1$, the contributions of far-away neurons could be as strong as the contributions of close-by neurons, especially if activity is correlated.

In more detail, let the ring be of inner radius $d$ and outer radius $d+\epsilon$, then its area is $A = 2 \pi (d \epsilon + \epsilon^2)$. At a constant density $\rho$, the number of neurons in this ring is given by $N=\rho A$, so that, up to a constant, $N \sim 2 \pi d \epsilon \rho$. Let $x_i = \{0, 1\}$ denote the state of a single neuron. Here, we assume that all neurons are uncorrelated and independent identically distributed random variables, with expectation value $\langle x \rangle = \mu$ and the same variance $\textrm{Var}[x]$. Our electrode potential was modeled as
\begin{equation}
V = \sum_{i=1}^N x_i /d_i\,.
\end{equation}
For every ring, we assume a constant $d_i = d$ for all neurons in the ring. Then, the expected potential for the ring is
\begin{equation}
\langle V(d) \rangle
= \left \langle \sum_{i=1}^N x_i /d \right \rangle
= \sum_{i=1}^N \langle x_i /d \rangle
= N \mu / d \,.
\end{equation}
Thus, when $N \sim 2 \pi d \epsilon \rho$, then $\langle V(d) \rangle \sim \mu$.
Indeed, this implies that the contributed potential of any of these rings is constant and that many neurons in a far-away ring can \enquote{contribute as much} as few local ones.

However, the variance of the potential per ring is not constant. For uncorrelated $x$, the variance of their sum is equal to the sum of their variances:
\begin{equation}
    \textrm{Var}[V(d)] = \textrm{Var}\left[\sum_{i=1}^N x_i /d \right] = \sum_{i=1}^N \textrm{Var}[x_i /d] \,.
\end{equation}
With
\begin{align}
    \textrm{Var}[x_i / d] &= \langle (x_i / d)^2 \rangle - \langle x_i / d \rangle^2 \\
    &= \frac{1}{d^2} (\langle x_i^2 \rangle - \langle x_i \rangle^2) \\
    &= \frac{1}{d^2}\textrm{Var}[x_i]
\end{align}
we see that
\begin{equation}
\textrm{Var}[V(d)] = \frac{1}{d^2}\sum_{i=1}^N \textrm{Var}[x_i] = \frac{N}{d^2} \textrm{Var}[x] \,.
\end{equation}
When again considering $N \sim 2 \pi d \epsilon \rho$,
\begin{equation}
\textrm{Var}[V(d)] \sim \textrm{Var}[x] \epsilon \rho / d \,.
\end{equation}
Hence, the standard deviation of the rings vanishes as $d \to \infty$; far away rings do contribute to the electrode, but they do not add to the variance of the signal. During the avalanche detection, the start or end of an avalanche is given by a threshold crossing. In a signal with more variance, more threshold crossing will occur, possibly leading to different avalanche statistics. As we showed, far-away neurons that are uncorrelated increase the mean, but they do  not increase the variance, and, thus, do not lead to more (or less) threshold crossings. Notwithstanding, this reasoning only holds for uncorrelated neurons. If $x_i$ are correlated, also far-away neurons could contribute to the variance.
Although we think that is not the main cause for the coarse-sampling effect (cf.~Fig~\ref{fig:supp_gamma_gaus_comparison}), we want to stress the limited range of electrodes for future work.

For a population of neurons receiving (un-) correlated synaptic input, the distance dependence of the LFP signal is studied in much more detail in Ref.~\cite{Linden2011}. Together with the above reasoning, this work highlights another possible source of sampling bias (which also affects our model electrodes):
Because closer-to-critical dynamic states typically feature more correlated activity than subcritical states, the effective distance-dependence of an electrode is also affected by the dynamic state that is being recorded.

To conclude this section, we \changed{want to sketch} an idealized experimental set-up:
in order to determine criticality under coarse-sampling, the set-up should combine i) a large distance between electrodes $d_{\rm E}$, ii) a narrow electrode field-of-view (large $\gamma$) and, ideally, iii) systems to calibrate with, which feature different dynamic states. This could potentially be used to qualitatively compare the distance to criticality between the systems. However, not only is this much more limited than what is possible with spike data~\cite{Wilting2018, Wilting2019, Levina2017}, but the cut-off is a characteristic and ubiquitous feature commonly observed in experimental data of coarse-sampled recordings~\cite{Beggs2003,Gireesh2008}. This indicates that electrodes typically have a large field-of-view, and motivated our modeling assumption of $\gamma=1$.

\subsection{Neuron density}

\begin{figure*}[!t]
\centering
\includegraphics[width=\textwidth]{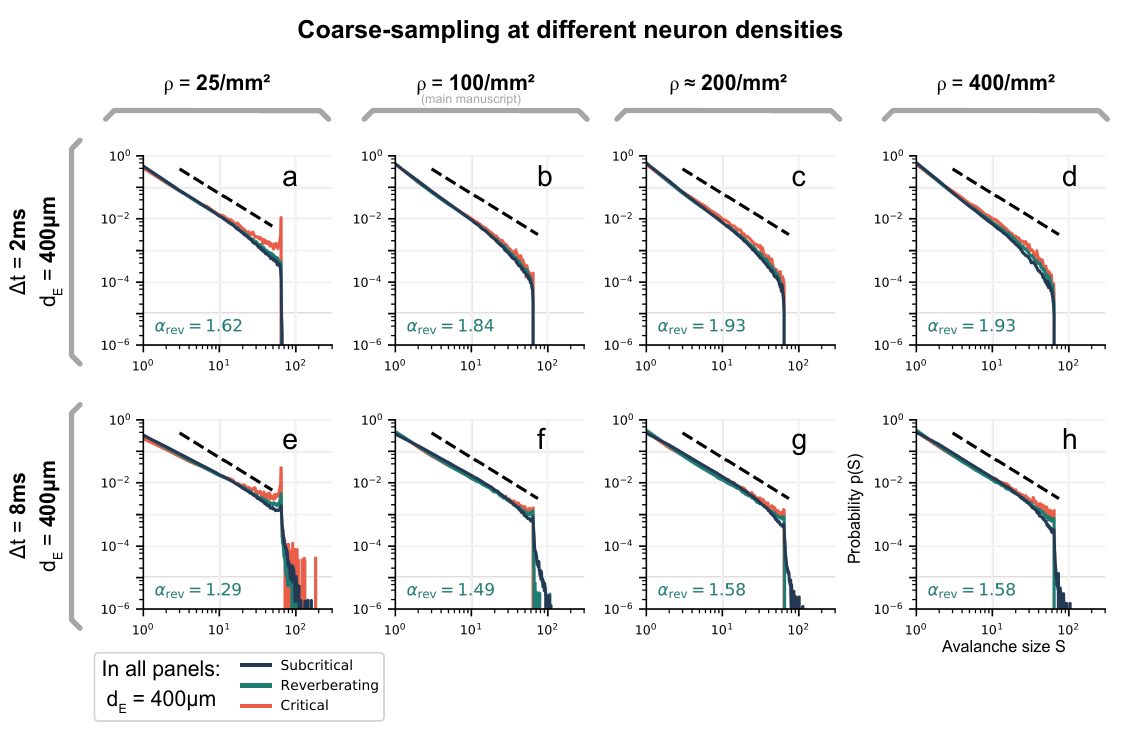}
\caption[Varying neuron densities]{{\bf Varying neuron densities.}
Simulations with parameters as in the main manuscript, with matching spatial extend of the substrate, connectivity profile of the neurons and field of view of the electrodes ($\gamma =1$).
As a guide to the eye, black dashed lines indicate $p(S) \sim S^{-1.5}$, and the fitted $\alpha$ for the reverberating state are indicated in the lower corner of every panel.
Our main result that measurement overlap can render dynamic states indistinguishable is confirmed.
}
\label{fig:supp_neuron_density}
\end{figure*}

As the coarse sampling effect is sensitive to the field of view of electrodes, it may similarly depend on the amount of neurons any electrode captures, and, thus, the density of the neuron population.
Hence, we performed a basic test of the robustness of our results with respect to the density of neurons (Fig~\ref{fig:supp_neuron_density}).
To that end, we kept most parameters of the model as in the main manuscript: The culture extended over $4 \cm$ substrate and neurons were hard-wired to their $\approx 1000$ closest neighbors. A change in density impacts how far these neighbors are distributed, but we kept the effective projection range constant ($\sigma = 300 \mum$), so that our \enquote{local} topology and the connectivity profile remain unchanged.
We kept electrode contributions at $1/d$ and placed electrodes at a large distance ($400 \mum$), so that every electrode samples many neurons and changes in density can become clear.

Compared to $\rho = 100 / \mm^2$ of the main manuscript, we considered lower ($\rho = 25 / \mm^2$) and higher densities ($\rho = 400 / \mm^2$).
Surprisingly, a change of neuron density only has a minor impact on the coarse sampling effect: Overall, the overlap of distributions and the cut-off remain for all considered densities (Fig~\ref{fig:supp_neuron_density}).
Nonetheless, subtle differences are visible.
Firstly, the slopes of the distributions seem to change, but this is shadowed by the dependence of the slope on the time-bin size.
Secondly, for higher densities $\rho \geq 100 / \mm^2$ distributions appear critical or slight subcritical, but for the low density, they also resemble super-critical distributions with a pronounced peak near the cut-off.

In conclusion, our main results concerning overlapping avalanche distributions for different dynamic states seem to be fairly invariant when increasing neuron density.

\subsection{Low-pass filtering}

\begin{figure*}[!t]
\centering
\includegraphics[width=\textwidth]{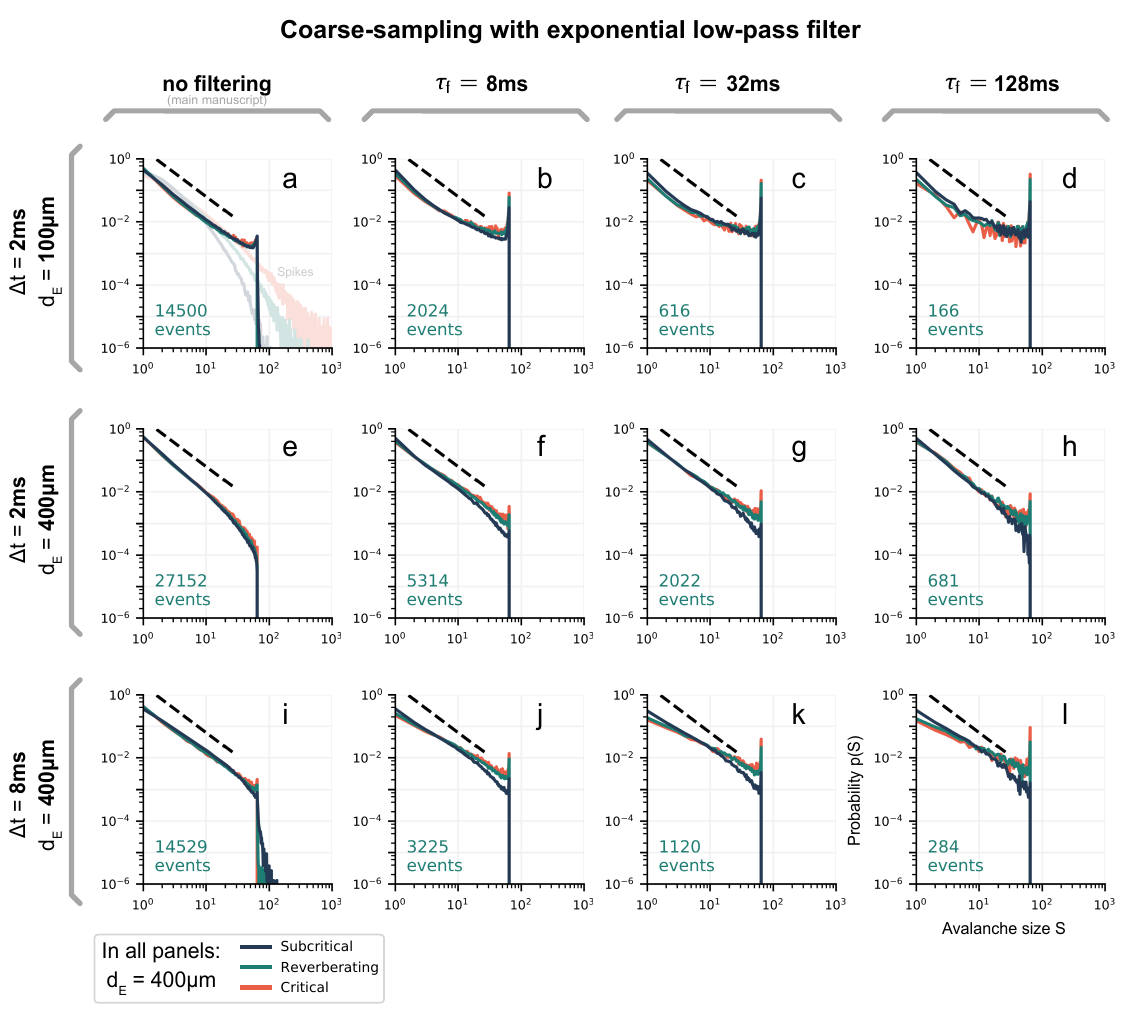}
\caption[Low-pass filtering using an exponential kernel]{{\bf Low-pass filtering using an exponential kernel.}
To mimic low-pass filtering as possibly caused by extracellular tissue or neuron morphology, an exponential kernel with decay time $\tau_{\rm f}$ was applied to the raw signal before applying the analysis pipeline of the main manuscript (which includes frequency filtering to $0.1 \Hz < f < 200 \Hz$).
As a guide to the eye, black dashed lines indicate $p(S) \sim S^{-1.5}$, and the number of events that contributed to the distribution for reverberating dynamics are indicated in the lower corner of every panel.
}
\label{fig:supp_filtering}
\end{figure*}

A relevant question that we have not addressed in the main manuscript is how mechanisms that come into effect \textit{before} the sampling hardware could impact avalanche statistics.
Many studies are concerned with low-pass frequency filtering and how measurements of neuronal activity are affected.
Such a temporal filtering may arise from the intrinsic neuronal morphology~\cite{Linden2011, Buzsaki2012, Einevoll2013}
or the surrounding extracellular tissue~\cite{Gabriel1996, Bedard2006, Bedard2009}.

As a simple test that mimics natural low-pass filtering, we convolved the raw time series of every electrode with an exponentially decaying kernel (Fig~\ref{fig:supp_filtering}).
Thus, the filtering was applied before the remaining analysis pipeline.
The decay of the kernel (and the strength of the filtering) is parametrized through the decay time of the exponential. We considered decay times $\tau_{\rm f}$ between $2 \msec$ and $128 \msec$ (ranging from $1$ to $128$ time steps of the simulation). The kernel was created using \texttt{scipy.signal.exponential} with a window size of $1000$ time steps.
The remaining analysis pipeline remained unchanged, and, in particular, included the frequency filtering to $0.1 \Hz < f < 200 \Hz$ that we assumed as part of the recording hardware.

Whereas the overlap of distributions largely remains when low-pass filtering is applied, the shape of the distributions depends on the strength of the filter.
As a general trend, all distributions tend to form \enquote{super-critical} peaks as filtering becomes stronger. We associated these peaks with multiple electrodes picking up the same event (boosting the amount of large avalanches, up to the number of electrodes).
This goes along with a decreased number of total avalanches that are detected (lower left corner in all panels). Note that the same raw time series with the same duration were used along every row of Fig~\ref{fig:supp_filtering}.

Together, this is consistent with the expecting \enquote{smoothing} of the raw time series due to low-pass filtering:
Deflections of a time series around its mean get attenuated, and small excursions (at high frequency) become rare. Because these excursions potentially trigger the start of a new avalanche, \changed{fewer} avalanches (or events) are detected when filtering becomes stronger. Intriguingly, the change of the distribution shape due to filtering seems to affect critical and reverberating dynamics more severely than subcritical dynamics (especially visible in the bottom row).

However, note that the filtering employed here only serves as an example for a low-pass filter. Experimentally, power-spectra are often found to show $1/f^\beta$ scaling, with $0 < \beta < 4$, which limits the functional form a more realistic filtering kernel might have~\cite{Bedard2006,Miller2009,Milstein2009,donoghue_parameterizing_2020}.


\subsection{Scaling laws may fail under coarse-sampling}

\begin{figure*}[t]
\centering
\includegraphics[width=\textwidth]{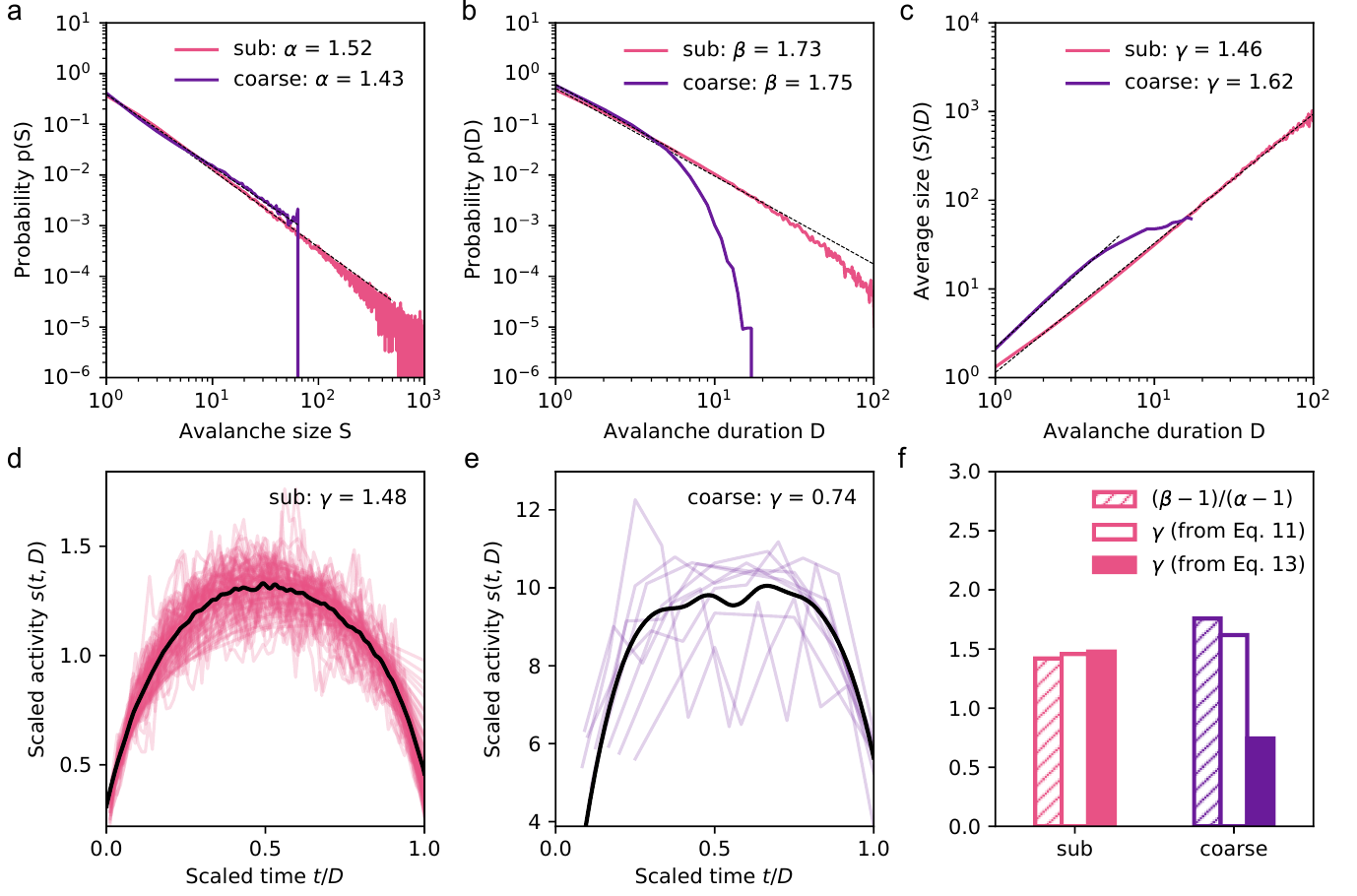}
\caption[Scaling laws of a system with critical dynamics under coarse- and sub-sampling]{\textbf{Scaling laws of a system with critical dynamics under coarse- and sub-sampling.}
{\bf a--c:}~Avalanche-size distribution $p(S)\sim S^{-\alpha}$,
avalanche-duration distribution $p(D)\sim D^{-\beta}$, and
average size for a given duration $\langle S \rangle (D) \sim D^\gamma$, respectively,
for sub-sampled (\enquote{sub}) and coarse-sampled (\enquote{coarse}) simulations.
Distributions under sub-sampling easily span more than one order of magnitude, while coarse-sampled distributions suffer from an early cut-off (which hinders power-law fits).
{\bf d, e:}~Shape collapse of $s(t,D) \sim D^{\gamma -1} \mathscr{F} (t/D)$ for sub-sampled and coarse-sampled data, respectively.
Under coarse-sampling, the early duration cut-off results in few unique shapes for the collapse (corresponding to unique $D$-values).
{\bf f:}~Comparison of the critical exponents obtained independently from Eqs.~\eqref{eq:p_avg}--\eqref{eq:shape_collapse}.
Exponents are consistent only under sub-sampling.
{\bf Parameters:}~$d_{\rm E} = 400 \mum$ and $\Delta t = 8 \msec.$
}
\label{fig:scaling}
\end{figure*}

The most used indication of criticality in neuronal dynamics is the avalanche-size distribution $p(S)$.
However, at criticality, the \textit{avalanche duration distribution} $p(D)$ and the \textit{average avalanche size} for a given duration, $\langle S \rangle (D)$, should also follow power-laws, each with a respective \textit{critical exponent}~\cite{Sethna2001}:
\begin{equation}
    p(S) \sim S^{-\alpha}
    \label{eq:p_s}
\end{equation}
\begin{equation}
    p(D) \sim D^{-\beta}
\end{equation}
\begin{equation}
    \langle S \rangle (D) \sim D^{\gamma}
    \label{eq:p_avg}
\end{equation}
The exponents are related to one another by the scaling relationship\footnote{
    \changedd{
        In this subsection, $\gamma$ exclusively denotes the scaling exponent and the decay exponent (which is denoted with $\gamma$ in the rest of the manuscript) equals 1 for all results presented here.
    }
}
\begin{equation}
    \frac{\beta-1}{\alpha-1} = \gamma\,.
    \label{eq:scaling}
\end{equation}
For a pure branching process \Ldash or any process in the mean-field directed percolation universality class \cite{Sethna2006, Janssen2005} \Rdash they take the values $\alpha=3/2$, $\beta=2$ and $\gamma=2$.

Lastly, at criticality, avalanches of vastly different duration still have the same \textit{average shape}:
The activity $s(t,D)$ at any given time $t$ (within the avalanche's lifetime $D$) is described by a universal scaling function~$\mathscr{F}$, so that
\begin{equation}
    s(t,D) \sim D^{\gamma -1} \mathscr{F} (t/D)\,.
     \label{eq:shape_collapse}
\end{equation}
In other words, changing $s(t,D)\rightarrow s(t,D)/D^{\gamma -1}$ and $t\rightarrow t/D$ should result in a data collapse for the average avalanche shapes of all durations.

The shape collapse of Eq. \ref{eq:shape_collapse} is done following the algorithm described in \cite{Marshall2016}. Briefly, the avalanche profiles $s(t,D)$ of all avalanches with the same duration $D$ are averaged, and the resulting curve is scaled to $t/D$ and interpolated on $1000$ points in the $[0,1]$ range. Avalanches with $D<4$ , or with less than $20$ realizations are removed. The chosen collapse exponent $\gamma$ is the one that minimizes the error function:
\begin{equation}
   E=\frac{\langle \text{Var}(X_D/D^{\gamma-1}) \rangle}{\Delta X^2}
    \label{eq:error_collapse}
\end{equation}
where $X_D(t/D)$ is the interpolated average shape of avalanches with size $D$, and $\Delta X = \max_{t,D} (X_D/D^{\gamma-1}) - \min_{t,D} (X_D/D^{\gamma-1})$. The variance $\text{Var}(.)$ is calculated over all valid $D$, and the mean $\langle . \rangle$ is taken over the scaled duration $t/D$. For interpolation and minimization we use the scipy \cite{Virtanen2020} functions \mbox{interpolate.InterpolatedUnivariateSpline} and \mbox{optimize.minimize}, respectively.

From Eqs.~\eqref{eq:p_avg}--\eqref{eq:shape_collapse}, we have three independent ways to determine the exponent $\gamma$.
Consistency between the three is a further test of criticality.
However, to the best of our knowledge, experimental evidence with the full set of scaling laws is mainly observed under sub-sampling: from spikes of in vitro recordings~\footnote{
An exception can be found in Ref.~\cite{Ponce-Alvarez2018}, where scaling relations were found to hold in vivo. However, in this study, fluorescence imaging was coupled with very large time bins $\Delta t \geq 0.47 \sec$, the effect of which remains to be understood in full.}~\cite{Friedman2012, Kanders2020}, but see Ref.~\cite{Ponce-Alvarez2018}.

The absence of scaling laws in coarse-sampled data can be explained by how coarse-sampling biases the average shape:
the cut-off in $p(S)$ near the number of electrodes $S = N_{\rm E}$ implies that $\langle S\rangle (D) < N_{\rm E}$.
From Eq.~\eqref{eq:p_avg} we have $D < N_{\rm E}^{1/\gamma}$.
If $\gamma > 1$ the cut-off in $p(S)$ causes a much earlier cut-off in both $p(D)$ and $\langle S\rangle (D)$.

\begin{figure}[thb!]
\centering
\includegraphics[width=7.0cm]{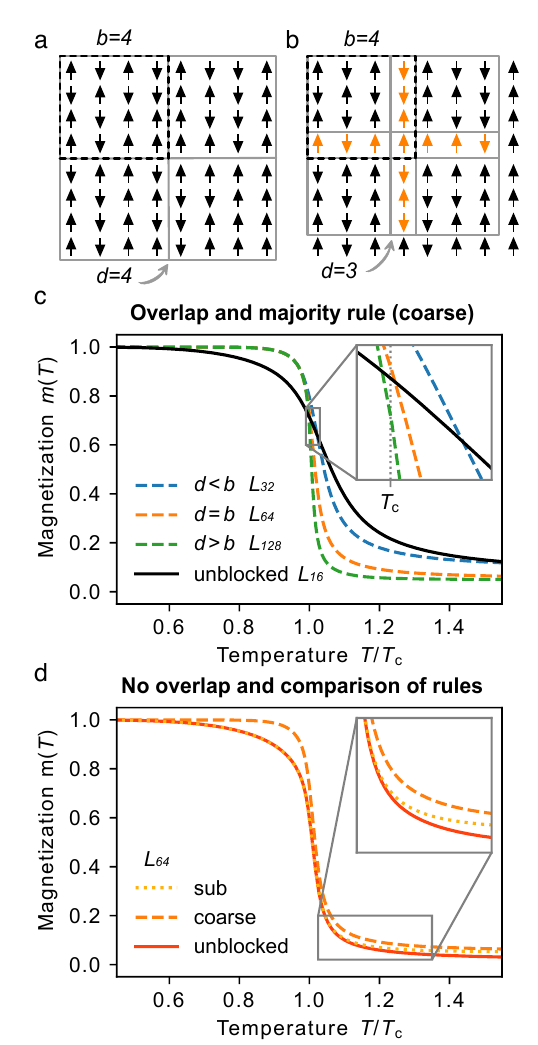}
\caption[Coarse graining the Ising model]{{\bf Coarse graining the Ising model.}
{\bf a:}~Representation of the standard coarse graining where block size matches the distance between blocks ($d=b=4$).
No overlap is created.
{\bf b:}~Coarse graining with block size $b=4$ and a distance between blocks of $d=3$.
Overlapping spins (orange) are shared by two or more blocks.
{\bf c:}~With the \enquote{coarse} majority rule, overlap impacts the spontaneous magnetization $m(T)$.
Only the crossing between the unblocked ($L=16$) and non-overlapping blocked system ($d=b$, $L=64$) happens at $T=T_{C}$, as would be expected.
Intriguingly, the overlap ($d<b$, $L=32$) pushes the system towards higher magnetization where spins appear more aligned.
On the other hand, the absence of overlap ($d>b$, $L=128$) causes smaller magnetization where spins appear more random.
(Note that, in order to avoid finite-size effects, the target size after coarse graining has to match, here $L=16$.
Consequently, depending on the ratio between $d$ and $b$, simulations have different system sizes.)
{\bf d:}~Comparison between the fully-sampled, unblocked system and blocked systems using the majority rule (\enquote{coarse}) and the decimation rule (\enquote{sub}) for $d = b = 4$. All simulations and curves for $L=64$.
In the ordered, low-temperature phase, the sub curve matches the fully sampled system.
Only for the high-temperature phase deviations occur due to finite-size effects (the magnetization for $T \to \infty$ approaches the value expected for the rescaled $L=16$ system).
The coarse curve is systematically biased towards more ordered states.
}
\label{fig:supp_ising}
\end{figure}

Given that experiments typically have $N_{\rm E} \sim 10^2$ electrodes, $p(D)$ of a pure branching process (with $\gamma=2$) would span a power-law for less than one order of magnitude. However, the typical standard to reliably fit a power-law is at least two orders of magnitude \cite{Stumpf2012}.
While this is problematic under coarse-sampling (Fig~\ref{fig:binsize}), we have shown that the hard cut-off is not present under sub-sampling (Fig~\ref{fig:binsize_sub}).


When we apply the independent measurements of $\gamma$ to our model (with critical dynamics) under sub-sampling, we find consistent exponents for all measurements (Fig~\ref{fig:scaling}f). Moreover, the exponents we find under sub-sampling are compatible with experimental values, e.g.~$\gamma_{\rm exp}=1.3\pm0.05$ reported in Ref.~\cite{Friedman2012} and $1.3 \leq \gamma_{\rm exp} \leq 1.5$ reported in Ref.~\cite{Kanders2020}.
Notably, the exponents found in our model and experimentally differ from those expected for a pure branching process ($\gamma = 2$). While not the focus here, we believe this deviation to derive from topological properties of the network, which was also observed in Ref.~\cite{Friedman2012}; distance-dependent weights of local topologies affect avalanche duration and size and yield different exponents than a branching process (which does not face any topological constraints).

When we apply the independent measurements of $\gamma$ to our model (with critical dynamics) under coarse-sampling, exponents differ from measurement to measurement:
The exponent obtained from the shape collapse
($\gamma \approx 0.74$)
greatly differs from the other two
($\gamma \approx 1.74$, $\gamma \approx 1.62$), Fig~\ref{fig:scaling}f.
Moreover, the extremely short range available to fit $p(D)$ and $\langle S \rangle (D)$ with power-laws ($1\leq D \leq 6$) makes the estimated exponents unreliable.

To conclude, the full set of critical exponents revealed criticality only under sub-sampling. Only in this case we observed both, a match between all the measurements of the exponent $\gamma$, and a power-law behavior extending over a range large enough to reliably fit them.

\subsection{Coarse Graining the Ising Model}

To demonstrate how general the impact of measurement overlap is, we study the two-dimensional Ising model. The Ising model is well understood and often serves as a text-book example for renormalization group (RG) theory in Statistical Physics~\cite{Newman1999}.
In this framework, the system is \textit{coarse grained} by merging multiple parts (spins) into one. An intuitive way to think of it is by zooming out of a photograph on a computer screen; a pixel can only show one color although there might be more details hidden underneath.
Coarse graining is also known as the real-space block-spin renormalization and it can be used to assess criticality.
Please note that \textit{coarse graining} is different from \textit{coarse-sampling}. Conventionally, coarse-graining perfectly tiles the space without any measurement-overlap (see Fig~\ref{fig:supp_ising}).

The two-dimensional Ising model consists of $N=L^{2}$ spins with states $s_{i}=\pm1$, arranged on a square lattice of length $L$.
In its simplest form, it is given by the Hamiltonian $H(\vec{s})=\sum_{\langle i,j\rangle}s_{i}s_{j}$, where $\langle i,j\rangle$ denotes all pairs of nearest neighboring spins.
The probability of observing $\vec{s}$ is given by the Boltzmann distribution
\begin{equation}
P(\vec{s},T)=\frac{1}{Z_{T}}e^{-H(\vec{s})/k_{B}T}
\end{equation}
where $T$ is the temperature of the system, $k_{\rm B}$ is the Boltzmann constant (here, $k_{\rm B}=1$) and $Z_{T}$ is the partition function that normalizes the distribution.
As the temperature $T\rightarrow T_{c}=2/\mbox{ln}(1+\sqrt{2})$, the system undergoes a second-order phase transition between a disordered spin configuration ($T>T_{c}$) and an ordered state of aligned spin orientations ($T<T_{c}$).
Many observables diverge at $T=T_{c}$ for $L\rightarrow\infty$, such as correlation length, specific heat and susceptibility~\cite{Newman1999,Sethna2006}.

We perform Monte Carlo simulations of the 2D Ising model using the massively parallel multicanonical method~\cite{Zierenberg2013,Gross2018}.
The multicanonical method offers numerous advantages over conventional Monte Carlo approaches.
For instance, instead of simulating at a single temperature, one simulation covers the whole energy space. High-precision canonical expectation values of observables are recovered for any desired temperature during a post-production step.
Thereby, we obtain the normalized absolute magnetization as a function of temperature $m(T)=\frac{1}{N}|\sum_{i}s_{i}|$.

\subsection{Block-Spin Transformation}
Measurement overlap causes individual sources to contribute multiple times to a signal. For the Ising model, a similar process takes place when coarse graining is applied.
In the process, spins are grouped into blocks of size $b\times b$, here $b=4$ and every block only takes a single value.
The value of each block can be obtained in different ways.
\begin{itemize}
\item
Most commonly, the majority rule~\cite{Newman1999} is employed, where the block is assigned $+1$ ($-1$) if the majority of spins has value $+1$ ($-1$).
In this case, the contribution of multiple sources is integrated. Hence we compare this rule to the effects observed when neuronal systems are coarse-sampled.
\item
Alternatively, one can use the decimation rule~\cite{Newman1999}.
In this case, all except a single spin value within a block are discarded. The block value is assigned from the single spin that is kept.
Hence we compare this rule to the effects observed when neuronal systems are sub-sampled.
\end{itemize}
This block-spin transformation rescales the number of spins by a factor of $1/b^{2}$, effectively reducing system size (which will cause finite-size effects). It is well known, that when studying the magnetization, the effective size of the compared systems (after rescaling) has to match.

\subsection{Overlap}
To mimic the measurement overlap, we now introduce an overlap between the blocks of the Ising model coarse graining (Fig~\ref{fig:supp_ising}).
In the native block-spin transformation, blocks do not overlap. Then, in terms of spins, the linear distance $d$ between two blocks matches the block size $b=d=4$ (Fig~\ref{fig:supp_ising}a).
When the distance between blocks is smaller than the block size, $d<b$ (Fig~\ref{fig:supp_ising}b), measurement overlap is created, while when $d>b$ parts of the system are not sampled.
Clearly, the changes that such an overlap will cause on rescaled observables should depend on the rule used to perform the block-spin transformation.

Here, we look at combinations of block size $b=4$ with distance between blocks of $d=2$, $d=4$ and $d=8$.
In order to preserve the effective system size ($L=16$), we thus perform simulations for $L=32$, $L=64$ and $L=128$, respectively.

Using the majority rule and no overlap \Ldash which is the default real-space renormalization-group approach \Rdash
the procedure moves $m$ away from $m\left(T_{c}\right)$ (Fig~\ref{fig:supp_ising}c, $d=b$):
For $T<T_{c}$, $m$ is increased;
For $T>T_{c}$, $m$ is decreased.
Ordinarily, $T_{c}$ can be obtained by finding the crossing of $m$ between an unblocked ($L=16$) and a blocked ($L=64$, $b=4$) system \Dash only at $T_{c}$ is the measured $m$ invariant under block rescaling transformations.

\subsection{Majority Rule \enquote{coarse}}

What is the impact of the overlap for the majority rule?
For increasing overlap ($d<b$), the crossing occurs at $T>T_{c}$ (Fig~\ref{fig:supp_ising}c).
This is because sharing spins increases the correlations between blocks (pairwise and higher-order), making it more likely that the rescaled spins point into the same direction.
In other words, it biases the measurement of $m$ towards order, increasing our estimated critical temperature.

For absent overlap ($d>b$), only every other block is measured.
This decorrelates the spins near the borders of each block and, therefore, decreases the correlation between blocks.
As a consequence, the spin orientation of the blocked system moves towards disorder, decreasing the measured magnetization $m$.

\subsection{Decimation Rule \enquote{sub}}
If instead of the majority rule the decimation rule is used, the blocking procedure does not alter the correlation between spins before and after the transformation (Fig~\ref{fig:supp_ising}d).
As a consequence, the magnetization remains unaltered in general.
However, in the disordered phase, we still notice a systematic deviation from the unblocked system (with $L=64$).
This deviation can be fully attributed to finite-size effects:
The distribution of realizable magnetizations in the disordered phase follows a Gaussian with mean zero and variance proportional to the (effective) number of spins.
Due to the definition of the magnetization with absolute value, the expectation value of the magnetization for $T\to\infty$ is determined by the (effective) system size.

As was the case when sub-sampling neuronal systems, the increase in correlation that ultimately leads to biased observables is caused by integrating weighted contributions from various sources.
This is not the case when the decimation rule is applied.
Note that the impact of different block-transformation rules on $m(T)$ will not hold for all other canonical observables such as the energy $E(T)$~\cite{Newman1999}.

\printbibliography[heading=subbibliography]

\begin{table*}[p]
\centering
\caption[Compilation of experimental findings of neuronal avalanches]{\textbf{Compilation of experimental findings of neuronal avalanches}~\cite{neto_criticality_2020}. For cultures, ``region'' corresponds to the brain region neurons were extracted from. Range of $p(S)$ was either given in the text, or estimated visually. Exponents with ``$\approx$'' are not explicitly fitted but instead visually compared. \changedd{Notably, most studies reporting power-law distributions in-vivo built on coarse-sampled measures, or imaging (in which case the effect of the low sampling-frequency on avalanche distributions is not clear). We found only one reference in the literature where in-vivo spiking data were used \textit{and} power-law distributions occurred, although over a very limited range~\cite{Hahn2017}.}}
\makebox[\linewidth]{
\begin{tabular}{@{}rcllcc@{}}
\toprule
authors     & technique & region               & system           & exponent $\alpha$          & range of $p(S)$ \\ \midrule
Beggs \emph{et al.}, 2003~\cite{Beggs2003}     & LFP       & cortex & culture  & $\approx1.5$                        & 1-60            \\
Gireesh \emph{et al.}, 2008~\cite{Gireesh2008}   & LFP       & cortex               & culture   & $1.53\pm0.02$ & 1-32            \\
Petermann \emph{et al.}, 2009~\cite{Petermann2009} & LFP       & cortex               & monkey           & $\approx1.5$                       & 1-32            \\
Klaus \emph{et al.}, 2011~\cite{Klaus2011}     & LFP       & cortex               & monkey           & $\approx1.5$                        & 1-96            \\
Priesemann \emph{et al.}, 2013~\cite{Priesemann2013}& LFP       & varied               & human            & $1.58\pm0.06$            & 1-50            \\
Shew \emph{et al.}, 2015~\cite{Shew2015}      & LFP       & cortex (visual)      & turtle & 1.6-2.6                    & 3-200           \\
Clawson \emph{et al.}, 2017~\cite{Clawson2017}   & LFP       & cortex (visual)       & turtle           & $1.8\pm0.3$   & 2-500           \\
Meisel \emph{et al.}, 2013~\cite{Meisel2013a}    & MEEG      & whole head           & human            & $\approx1.5$                        & 1-27            \\
Palva \emph{et al.}, 2013~\cite{Palva2013}     & MEEG      & whole head           & human            & 1.31                       & 1-66            \\
Shriki \emph{et al.}, 2013~\cite{Shriki2013}    & MEEG      & whole head           & human            & $\approx1.5$                        & 1-100           \\
Arviv \emph{et al.}, 2015~\cite{Arviv2015}     & MEEG      & whole head           & human            & 1.50                       & 1-100           \\
Scott \emph{et al.}, 2014~\cite{Scott2014}     & imaging   & cortex               & rat              & $\approx1.5$                         & 1-1000          \\
Bellay \emph{et al.}, 2015~\cite{Bellay2015}    & imaging   & cortex               & rat              & $1.63\pm0.13$ & 0.01-10         \\
Fagerholm \emph{et al.}, 2015~\cite{Fagerholm2015} & imaging   & cortex               & rat              & $\approx1.5$                         & 1-1000          \\
Yaghoubi \emph{et al.}, 2018~\cite{Yaghoubi2018a}  & imaging   & hippocampus          & culture   & $1.65\pm0.1$  & 1-500           \\
Ponce-Alvarez \emph{et al.}, 2018~\cite{Ponce-Alvarez2018} & imaging & whole brain & zebrafish & $2.01\pm0.03$ & 1-10000\\
Bocaccio \emph{et al.}, 2019~\cite{Bocaccio2019} & imaging & whole head & human & $\approx2$  & 1-1000 \\
Friedman \emph{et al.}, 2012~\cite{Friedman2012}   & spikes     & cortex               & culture   & $1.7$   & 1-30            \\
Massobrio \emph{et al.}, 2015~\cite{Massobrio2015b} & spikes     & cortex               & culture   & 1.48-1.52                  & 1-500           \\
Hahn \emph{et al.}, 2017~\cite{Hahn2017}      & spikes     & cortex (visual)       & monkey           & $1.58\pm0.03$ & 1-20            \\
Levina \emph{et al.}, 2017~\cite{Levina2017} & spikes & cortex & culture & $\approx2$  & 1-5900\\
Kanders \emph{et al.}, 2020~\cite{Kanders2020} & spikes & hippocampus & culture & $2.18\pm0.05$ & 1-60\\
\bottomrule
\end{tabular}
}
\label{tab:ch2_experiments}
\end{table*}

\begin{figure*}[p]
\centering
\includegraphics{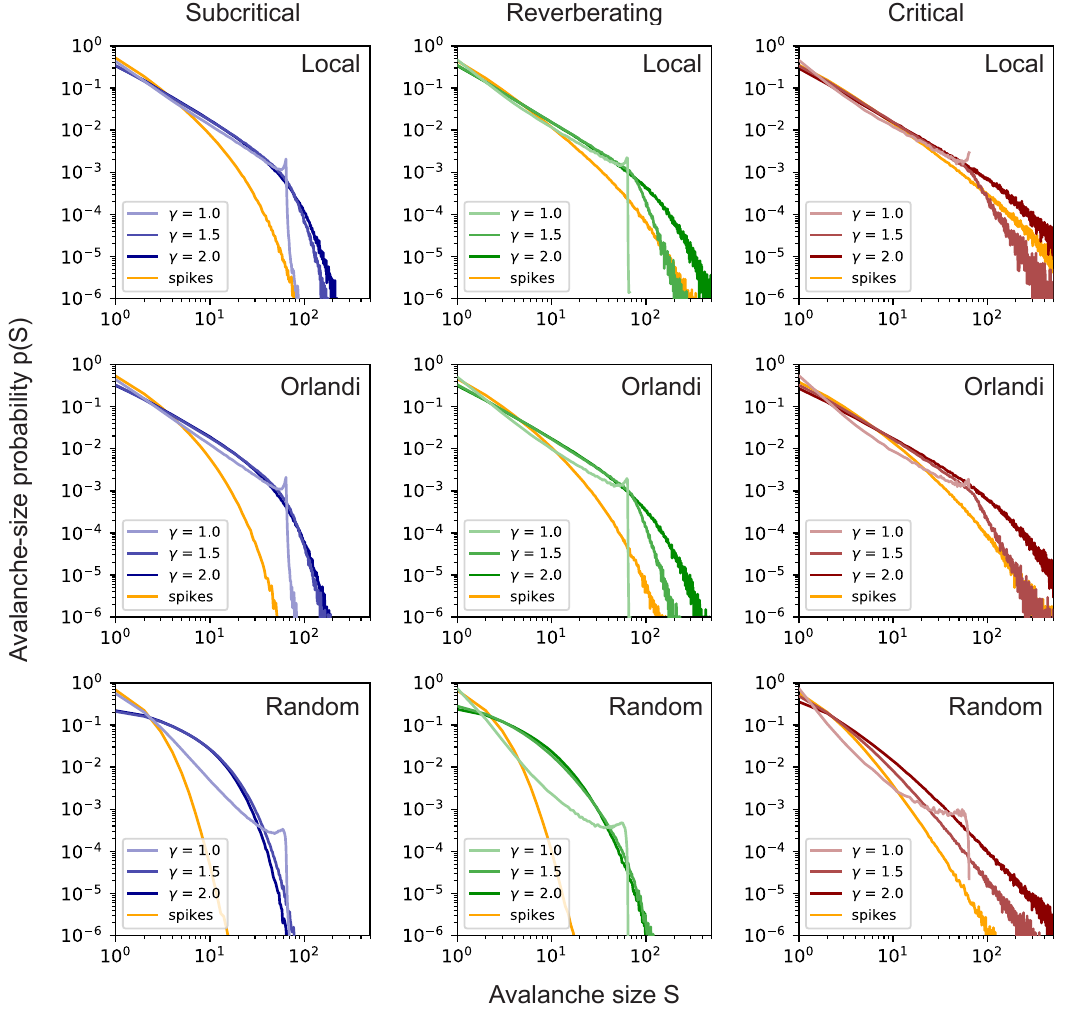}
\caption[Effect of changing the electrode contribution $\sim 1/d^{\gamma} $ of a spiking neuron at distance $d$, for different network topologies and $d_{\rm E}  = 200 \mum$]{{\bf Effect of changing the electrode contribution $\sim 1/d^{\gamma} $ of a spiking neuron at distance $d$, for different network topologies and $d_{\rm E}  = 200 \mum$.}
Dynamic states are Subcritical ({\bf left}), Reverberating ({\bf center}) and Critical ({\bf right}).
Topologies are Local ({\bf top}), Orlandi ({\bf middle}) and Random ({\bf bottom}). Local corresponds to the topology used in the main paper, Orlandi corresponds to the model described in \cite{Orlandi2013}, and Random corresponds to a completely random topology. Increasing $\gamma$ (decreasing electrode FOV) results in a loss of the cut-off for $p(S) \sim N_{\rm E}$ as the coarse-sampling becomes more spike-like. Bin-size for all distributions is $\Delta t = 4 \msec$.
}
\label{fig:supp_gamma_d08}
\end{figure*}

\begin{figure*}[p]
\centering
\includegraphics{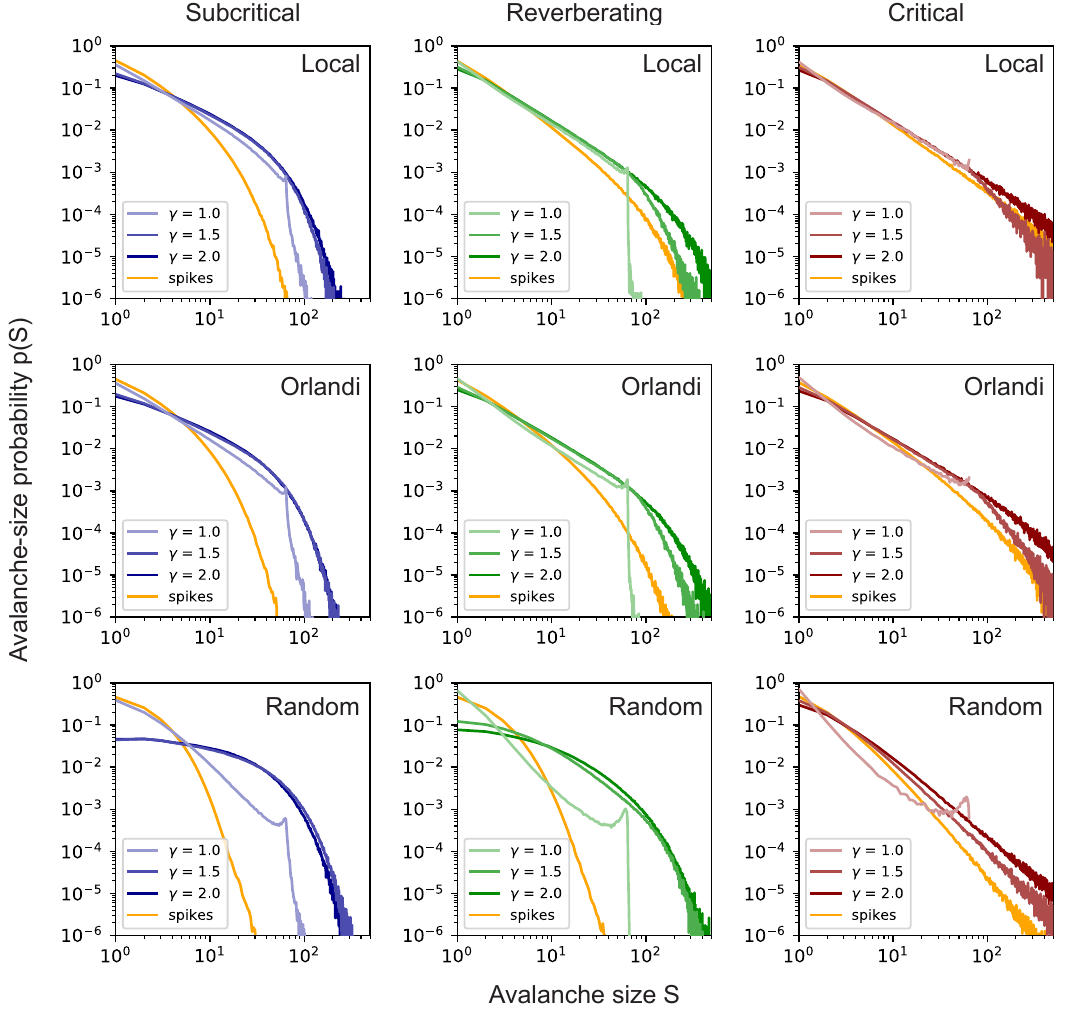}
\caption[Effect of changing the electrode contribution $\sim 1/d^{\gamma} $ of a spiking neuron at distance $d$, for different network topologies and $d_{\rm E}  = 400 \mum$]{{\bf Effect of changing the electrode contribution $\sim 1/d^{\gamma} $ of a spiking neuron at distance $d$, for different network topologies and $d_{\rm E}  = 400 \mum$.}
Dynamic states are Subcritical ({\bf left}), Reverberating ({\bf center}) and Critical ({\bf right}).
Topologies are Local ({\bf top}), Orlandi ({\bf middle}) and Random ({\bf bottom}). Local corresponds to the topology used in the main paper, Orlandi corresponds to the model described in \cite{Orlandi2013}, and Random corresponds to a completely random topology. Increasing $\gamma$ (decreasing electrode FOV) results in a loss of the cut-off for $p(S) \sim N_{\rm E}$ as the coarse-sampling becomes more spike-like. Bin-size for all distributions is $\Delta t = 8 \msec$.
}
\label{fig:supp_gamma_d04}
\end{figure*}

\begin{figure*}[p]
\centering
\includegraphics{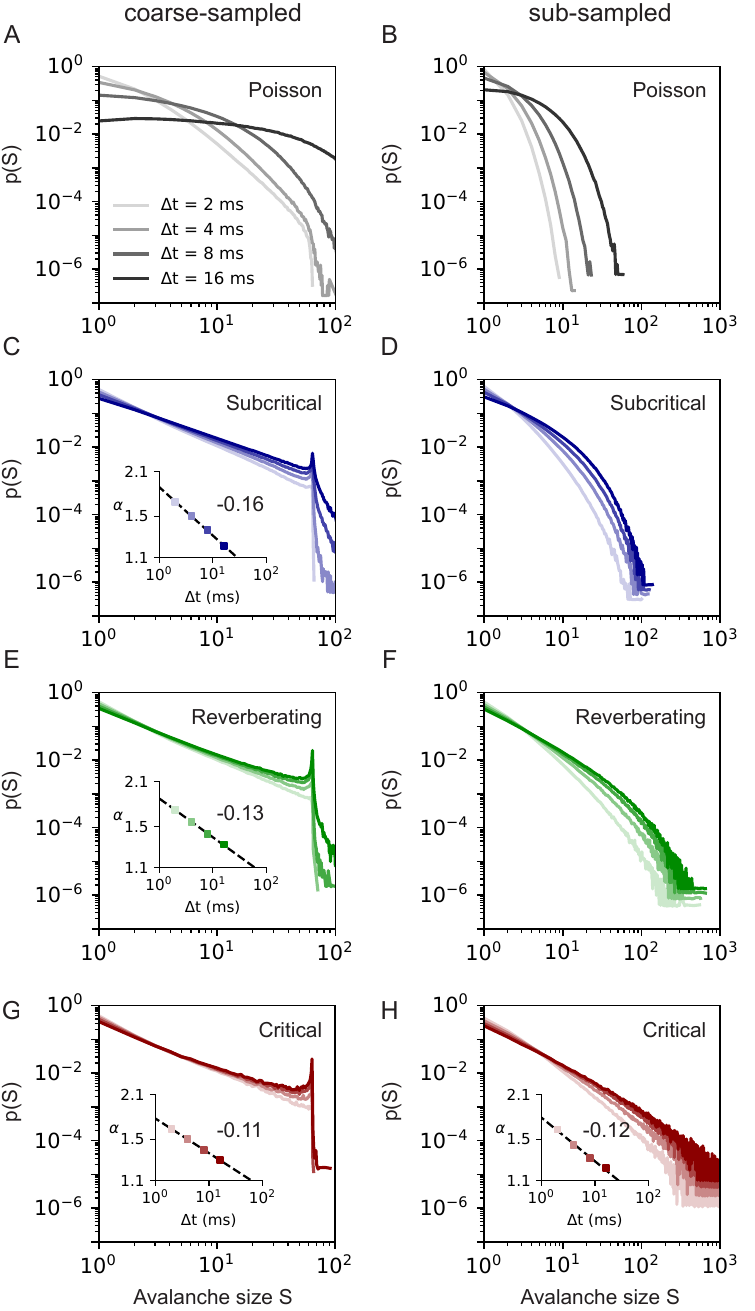}
\caption[Avalanche-size distributions $p(S)$ dependence on time-bin size $\Delta t$ for $d_{\rm E} = 200\mum$]{{\bf Avalanche-size distributions $p(S)$ dependence on time-bin size $\Delta t$ for $d_{\rm E} = 200\mum$.}
Coarse-sampled ({\bf left}) and sub-sampled ({\bf right}) results from an array of $64$ virtual electrodes with time bin sizes between $2 \msec \leq \Delta t \leq 16 \msec$. Dynamics states are Poisson ({\bf a-b}), Subcritical ({\bf c-d}), Reverberating ({\bf e-f}) and Critical ({\bf g-h}). Distributions are fitted to $p(S) \sim S^{\alpha}$.
{\bf Insets:}  Dependence of $\alpha$ on $\Delta t$, fitted as $\alpha \sim \Delta t^{-\beta}$. Fit values are shown in Table.~\ref{tab:fit_exponents}.}
    \label{fig:supp_binsize}
\end{figure*}


\end{refsection}

\end{document}